\documentclass[12pt,a4paper]{article}

\usepackage[margin=2.5cm]{geometry}

\usepackage{setspace}
\usepackage{graphicx}
\usepackage{amsmath, amssymb}
\usepackage{multirow}
\usepackage{pdflscape}
\usepackage[flushleft]{threeparttable}
\usepackage{tablefootnote}
\usepackage{color,soul}
\usepackage{natbib}
\usepackage[title]{appendix}
\usepackage{color, colortbl}
\usepackage{float}

\definecolor{Gray}{gray}{0.9}
\usepackage{subfigure}

\usepackage[table]{xcolor}
\usepackage{array}
\usepackage{makecell}
\usepackage{caption}

\usepackage{algorithm}
\usepackage[noend]{algpseudocode}
\usepackage{authblk}
\usepackage{hyperref}
\usepackage{eurosym}

\usepackage{tikz}
\usetikzlibrary{decorations.pathreplacing}
\usetikzlibrary{shapes.geometric}
\usetikzlibrary{shapes.multipart}
\usetikzlibrary{arrows,topaths}

\newcommand{\boldtheta}{\mbox{\boldmath{$\theta$}}}

\newcommand{\bV}{\mathbf{V}}

\newcommand{\bZ}{\mathbf{Z}}

\newcommand \deriv[2]  { \frac{\mathrm d{#1}}{\mathrm d{\,#2}}   }   
\newcommand \tausub[1] {\tau_{\mbox{\tiny #1}}}
\newcommand \fsub[1] {p_{\mbox{\tiny #1}}}
\newcommand \isuper[1] {I^{\mbox{\tiny #1}}}

\begin{document}
	
		\title{\vspace{-1cm}An age-structured SEIR model for COVID--19 incidence in Dublin, Ireland with framework for evaluating health intervention cost}

		\author[1]{\normalsize Fatima-Zahra Jaouimaa\footnote{These authors contributed equally to this work and are joint first authors of this article.}}
		\author[2]{\normalsize Daniel Dempsey$^*$}
		\author[3]{\normalsize  Suzanne van Osch}
		\author[3]{\normalsize Stephen Kinsella}
		\author[1]{\normalsize Kevin Burke\footnote{These authors also contributed equally to this work and are joint senior authors.}}
		\author[2]{\normalsize Jason Wyse{$^{\dag}$}}
		\author[1]{\normalsize James Sweeney{$^\dag$}\footnote{james.a.sweeney@ul.ie
		}}
	
		\affil[1]{\footnotesize Department of Mathematics \& Statistics, University of Limerick, Limerick, Ireland}
		\affil[2]{\footnotesize  School of Computer Science \& Statistics, Trinity College Dublin, Dublin, Ireland}
		\affil[3]{\footnotesize Kemmy Business School, University of Limerick, Limerick, Ireland\vspace{-0.5em}}
\date{\today\vspace{-1.5em}}

\vspace{-1.5em}
		\maketitle
		\begin{abstract}
					
		Strategies adopted globally to mitigate the threat of COVID--19 have primarily involved lockdown measures with substantial economic and social costs with varying degrees of success.  Morbidity patterns of COVID--19 variants have a strong association with age, while restrictive lockdown measures have association with negative mental health outcomes in some age groups. Reduced economic prospects may also afflict some age cohorts more than others. Motivated by this, we propose a model to describe COVID--19 community spread incorporating the role of age-specific social interactions. Through a flexible parameterisation of an age-structured deterministic Susceptible Exposed Infectious Removed (SEIR) model, we provide a means for characterising different forms of lockdown which may impact specific age groups differently. Social interactions are represented through age group to age group contact matrices, which can be trained using available data and are thus locally adapted. This framework is easy to interpret and suitable for describing counterfactual scenarios, which could assist policy makers with regard to minimising morbidity balanced with the costs of prospective suppression strategies. Our work originates from an Irish context and we use disease monitoring data from February 29th 2020 to January 31st 2021 gathered by Irish governmental agencies. We demonstrate how Irish lockdown scenarios can be constructed using the proposed model formulation and show results of retrospective fitting to incidence rates and forward planning with relevant ``what if / instead of'' lockdown counterfactuals. Uncertainty quantification for the predictive approaches is described. Our formulation is agnostic to a specific locale, in that lockdown strategies in other regions can be straightforwardly encoded using this model. The methods we describe are made publicly available online through an accessible and easy to use web interface. 	
	\end{abstract}
	
	\section{Introduction}

The global race to manage the existential threat posed by COVID--19 has used non-pharmaceutical lockdown (or restriction of movement) measures as a central tenet. These will continue to play a major role in public health policy as new variants emerge and before a full vaccine roll-out has been reached. As nations have come to terms with COVID--19 throughout the past year, the societal and economic impacts of the pandemic have become clear. A systemic shock to the world of work, widespread job losses in certain economic sectors, and a vast reduction in person-to-person social contact has given rise to an epoch of uncertainty, anxiety and fear. While older individuals are observed to be gravely threatened by the risk of infection, younger people have been particularly impacted by deteriorating mental health during this time \citep{Kwong20} in addition to reduced economic prospects \citep{Darmody20}. Governments have found themselves performing a difficult balancing act. Strict lockdown measures are necessary for public safety and to prevent health systems becoming overwhelmed. However, periods of strict measures need to be punctuated by temporary easing of restrictions whenever possible to give hope to businesses and reduce the psycho-social demands placed on citizens. National ``maps'' and ``road-plans'' for emerging from COVID--19 that were proposed in the first quarter of 2020 by national Governments have been tweaked and revised world-wide; the time elapsed since March 2020 has been characterised by an ebb and flow of various forms of restriction of movement, both within and between nations. 
\par
It seems that lockdown measures and their consequences will be present in citizens' lives for some time to come. Certain measures or guidelines to citizens may target specific age groups. For example, guidance has often urged extra protection for the elderly; in Ireland and the UK, this has been termed ``cocooning''. There has been much debate about the risks posed by keeping schools for children open and as a result there has been variation in school closures globally. Quantifying the potential impact of new or changing measures that target age cohorts differently is thus essential. The exploration and consideration of counterfactuals can provide valuable lessons and insights to policy makers at a time of much uncertainty.
\par
Our contribution in this article is to propose and calibrate a flexible model within the Susceptible Exposed Infected Recovered (SEIR) class, which can characterise different forms of lockdown measures with age structuring. The implied mortality burden of a lockdown architecture can be assessed through forward projection from the model. We show how this age-structured SEIR model, which includes explicit modelling of social mixing, can be used to assess and quantify the overall potential disease burden resulting from restriction of movement measures. Social mixing is modelled using age group to age group contact rates, allowing for assessment of the long run impact brought about by lockdowns which implicitly target specific age groups. The primary benefit of this approach is the potential to evaluate the `cost' of specific intervention actions in terms of impacts on the general public. If each of the public health interventions that are being considered can be economically costed, then a strategy for disease suppression in tandem with minimising economic costs can be explored. As an example, forward projection can indicate costs to economic sectors such as the night--time economy which relies heavily on young adults. This view is not constrained to economic costs alone, as public health authorities may alternatively focus on health costs such as mental health impacts brought about by constraints on social mixing amongst the general public. 
\par 
There has been some exploration of age-structured SEIR models for disease incidence. \citet{teimouri2020} constructed an age-structured SEIR model for the London area which incorporates contact tracing; their interest was detailing the impact of social mixing and contact tracing on the effective reproduction rate of the disease as opposed to model calibration. In a similar vein, \citet{grimm2021} use assumed epidemiological parameters to simulate the impact of age-specific control measures and contact tracing impact with a focus on the impact of control measures on factors including hospitalisations and deaths. \citet{lee2021} simulate the impact of four control measures, namely, school closure, social distancing, quarantine, and isolation on reproduction rates in South Korea using an age-structured SEIR model of disease spread. Maximum likelihood estimation is used to estimate contact scaling parameters, however no uncertainty in estimates or projections is presented. \citet{cuevas-maraver2021} propose a two-cohort age-segmented model (age in years $\leq$65/$>$65) for Mexican incidence counts and illustrate that age specific control measures may have utility for public health policy decisions. \citet{kimathi2021} simulate the impact of three specific governmental interventions on case incidence using an age-structured SEIR model with predetermined model parameters.

\par
The dynamics of the age-structured SEIR model we propose have a number of advantages over these competing approaches. We account for the impact of movement restrictions on population mixing by scaling age-structured contact matrices, as with \citet{prem2020}, however our scaling parameters are calibrated using the time series of observed Irish incidence counts as opposed to best-guess estimates. Where the parameters governing dynamics of models cannot be estimated due to data sparsity, we use recent results published in the COVID--19 literature on infection dynamics as well as expert opinion from the Irish Epidemiological Modelling Advisory Group (IEMAG); note that IEMAG developed an initial SEIR model \citep{iemag2020} that we extend. We use a statistical bootstrapping approach to present uncertainties in learned parameters, in addition to providing uncertainty intervals for incidence projections. 
\par
The available data for calibration of social contacts consists of daily count incidences and the specific lockdown measures implemented within Ireland from February 29th 2020 to January 31st 2021. While we present an analysis specific to an Irish context, we argue that the proposed approach is adaptable to other locales, wherein region specific macro-level behaviours can be calibrated. Furthermore, the framework we present can be easily adapted to incorporate more in depth population mixing knowledge from contact-tracing initiatives as well as allowing for estimation of all unknown model parameters.
\par
Based on the SEIR model described in this article, we have developed an R Shiny app interface that, given a user specified lockdown regime, creates an 8 week ahead forecast for estimated deaths and economic costs in Dublin. The app is accessible at \url{http://54.75.10.2:3838/ForecastApp/} with the app code available at \url{https://github.com/fatimZJ/Covid-19-Project}. 
\par
The remainder of the paper is organized as follows. Section~\ref{sec:data} provides an overview of observed COVID--19 incidence data and lockdown architectures in Ireland, a model of the effects of these lockdowns on social mixing by rescaling the population contact matrix over time, and the associated economic impacts. In Section~\ref{sec::epi_model} we present our proposed age-structured SEIR model which incorporates the scaled elements of the contact matrix, and, hence, lockdown effects. Section~\ref{sec::model_fit} describes calibration of the free parameters in the model, and quantification of uncertainty using a bootstrapping approach. Section \ref{sec::results} outlines the results of retrospective model fitting to Irish data, with associated projections and examples of some interesting counterfactual situations provided in Section \ref{sec:fwdcounter}. We conclude in Section \ref{sec::discussion} with a discussion. 

\section{Data sources} \label{sec:data}

We restrict our data sources to those that are typically freely publicly available, allowing for ease of implementation in other regions. The available data in an Irish context consists of daily incidence counts and the dates of changes of lockdown restrictions. Estimated contact matrices for age structured population mixing are sourced from literature. We defer discussion of the mechanistic parameters sourced from the COVID--19 literature to Section~\ref{sec::epi_model}.

\subsection{Daily incidence counts}

\begin{figure}[h]
	\centering
	\includegraphics[scale=0.75]{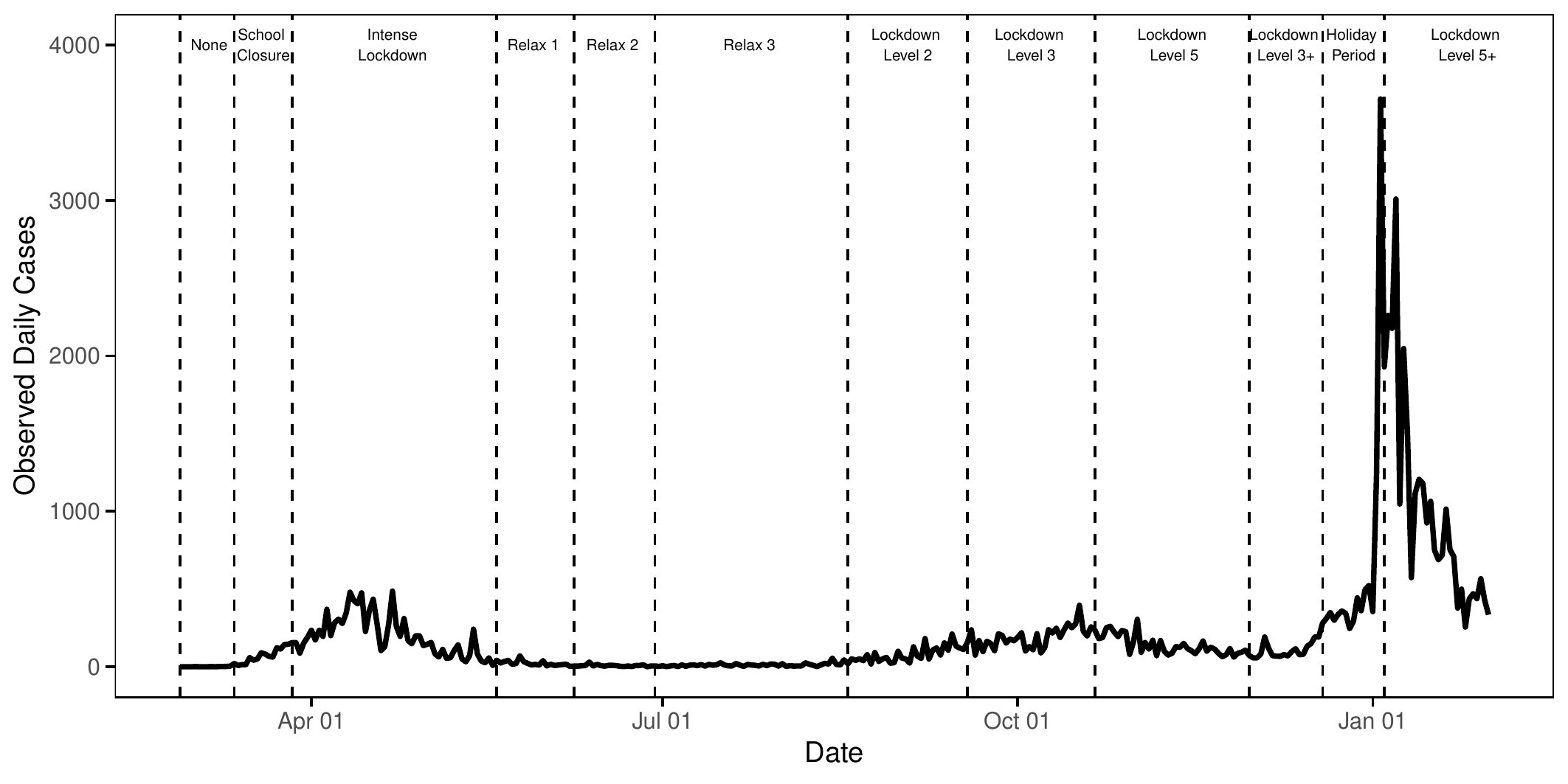}
	\caption{COVID--19 daily case incidence with corresponding lockdown levels in each period between February 2020 and January 2021. Descriptions of individual lockdown levels are presented in Table~\ref{tab::policy_summary}.}
	\label{fig::dailycases}
\end{figure}

The Irish Health Surveillance Protection Centre (\texttt{data.gov.ie}) provide anonymised daily COVID--19 incidences. We use the data from the period of February 29th 2020 to January 31st 2021 for model calibration. The daily COVID--19 count incidence is shown in Fig~\ref{fig::dailycases}. Age structured case count data is not publicly available in Ireland, and hence we use aggregate case counts at the population level. We use projected population data for 2019 provided by Irish Central Statistics Office (CSO) (\texttt{https://data.cso.ie/table/PEB07}) to estimate the age-structured population breakdown by county to estimate Dublin's population. Given the constraints on public movement, we make the assumption that the 2019 projections are representative of the population since the beginning of the pandemic.   

\subsection{Form of lockdown restrictions}

\begin{table}
	\footnotesize
	\centering
	\begin{tabular}{|c||c|c|c|c|c|}
		\hline
		&\textbf{Level 1}&\textbf{Level 2}& \textbf{Level 3} & \textbf{Level 4} & \textbf{Level 5} \\
		\hline
		\hline
		
		\textbf{House visits}  & 10  & 6  & 6 & 0 & 0\\
		(households)& (3) & (3) & (1) & \phantom{1} & \phantom{1 household}  \\
		\hline
		\textbf{Gatherings} &   & 6 indoor & 0 & 0 & 0 \\
		& 50 outdoor & 15 outdoor &  &  &  \\
		\hline
		\textbf{Weddings} & 100 & 50 & 25 & 6 & 6 \\
		& & & & & \\
		\hline
		\textbf{Indoor events} & 100 & 50 & 0  & 0 & 0 \\
		& & & & & \\
		\hline
		\textbf{Sporting events} & 100 indoor & 50 indoor & 0 & 0 & 0 \\
		& 200 outdoor & 100 outdoor &  &  &  \\
		\hline
		\textbf{Food venues} & Open & 6 & 15 & 15 & 0 \\
		& & (3 households) & outdoor & outdoor &  \\
		\hline
		\textbf{Pubs} & Open & 6  & 15 & 15 & 0 \\
		& & (3 households) & outdoor & outdoor &  \\
		\hline
		\textbf{Public transport} & 100\% & 50\% & 50\% & 25\% & 25\% \\
		\textbf{capacity} & &  &  &  &  \\
		\hline
	\end{tabular}
	\caption{Summary overview of the restrictions impacting on public gatherings for each of the five lockdown levels in Ireland. The numbers comprise the limits on individuals allowed to gather together in each social setting unless otherwise specified as a household limit. Details on other restrictions, such as on private travel, have been omitted for brevity.}
	\label{tab:lockdown_levels}
\end{table}
\par
Fig~\ref{fig::dailycases} shows the timeline and duration of varying degrees of restriction measures (vertical dashed lines) implemented in Ireland from March 2020. In line with \citet{iemag2020}, we define the 28th February 2020 as ``day zero'' of the Irish epidemic. As with many other countries, the Irish government introduced a strict lockdown in the early stages of the pandemic which lasted until May 2020. This lockdown was followed by a gradual easing of restrictions throughout the summer until case numbers began to rise in early autumn, when harsher restrictions were reintroduced and another phase of a strict lockdown was announced for late October. Restrictions were eased over the month of December but a subsequent wave of cases forced the implementation of a further strict lockdown immediately after the December holiday period. A more detailed overview of the restrictions is provided in Table~\ref{tab::policy_summary} in supplementary material A. 
\par
The nature of restrictions on public mobility in  Ireland, announced by the Irish government in April 2020, follow five levels. Level one is the least restrictive with this increasing to most restrictive at level five. In level one, food venues and bars remain open, gatherings of up to fifty people are permitted outdoors and sporting events can take place with restrictions on numbers. Level five corresponds to a near total blanket close on all activities. As the public health situation evolved during 2020, small adjustments were made to these levels with slight easing of targeted restrictions (for example, reopening of schools or childcare) within more severe lockdowns. An overview of the five level lockdowns is provided in Table~\ref{tab:lockdown_levels} with in-depth detail available at \url{gov.ie/en/campaigns/resilience-recovery-2020-2021-plan-for-living-with-covid-19/}. We denote the time intervals of lockdown measures using $\mathcal{I}_k = (r_{k},r_{k+1}]$, where $r_k$ is the time of the beginning of the $k$th regime for $k=1,\dots,N$ where $N=12$, with the first regime corresponding to no intervention from 29th February to 11th March 2020. Thus, we define $r_1 = 0$ corresponding to 29th February 2020 and $r_{N+1} = 336$ corresponding to 31th January 2021.
\subsection{Age structuring and social mixing\label{sec:agesocial}} 

COVID--19 is an airborne virus, hence consideration of close social mixing in the population is essential to capturing the observed patterns of infection. 
Furthermore, the strong association of morbidity patterns with the elderly, and more recently younger persons \citep{Taylorn879}, suggest consideration of age structured social mixing will be a key component of future projections~\citep{prem2020, cuevas-maraver2021}. Age-structured social mixing is typically captured in SEIR models through the use of age group to age group contact matrices. Although such matrices cannot capture the granular complexity of individual human interactions, they provide a reasonable approximation that can be incorporated into mathematical models for infectious diseases as demonstrated by~\citet{mossong2008}, and within this article. We follow~\citet{prem2020} and stratify the population into five-year bands from age 0 up to age 75, with one category for all individuals aged 75 and above, giving $A=16$ age groups.
\par
\begin{figure}[h]
	\begin{center}
		\includegraphics[scale=0.65]{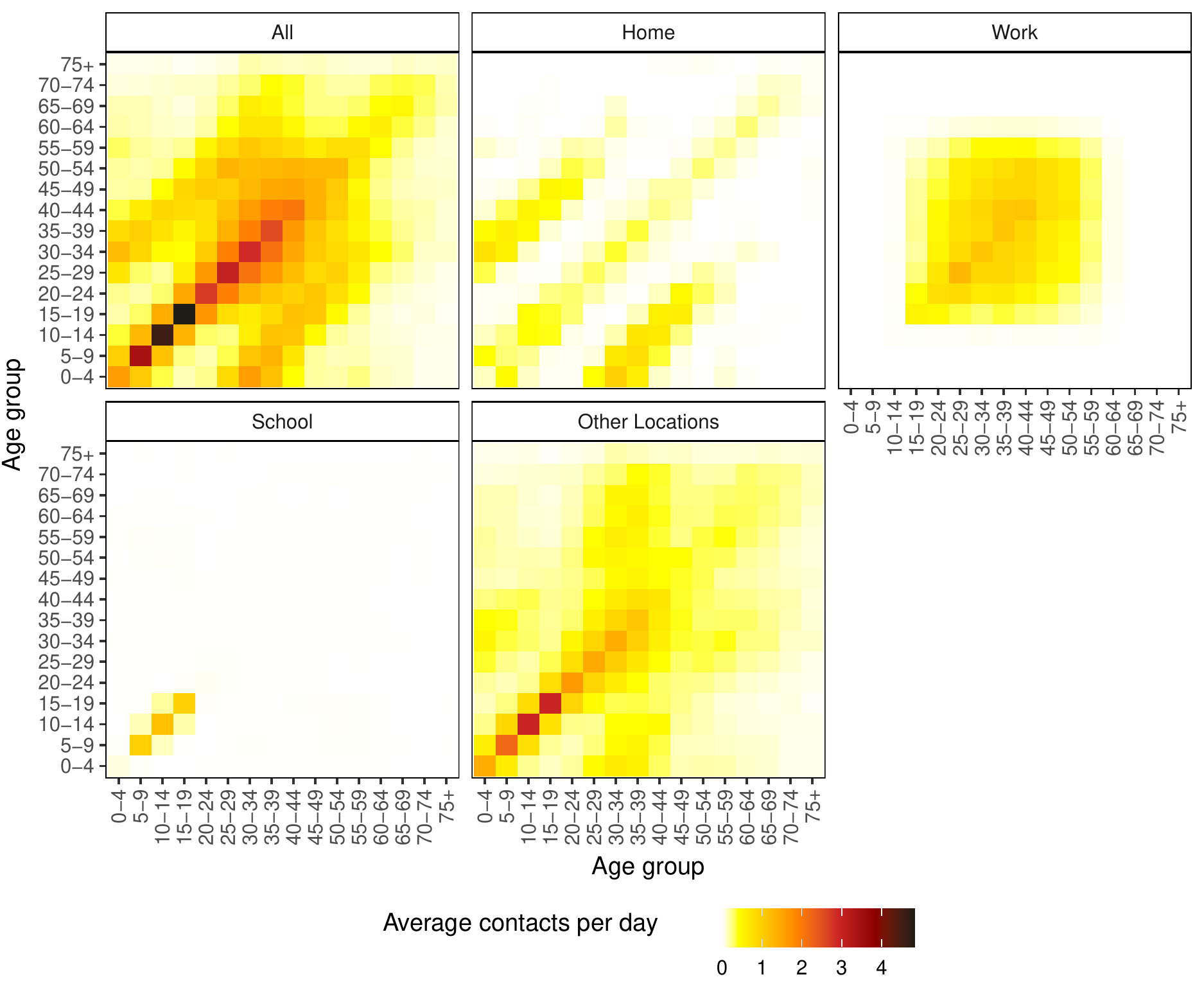}
	\end{center}
	\caption{Estimated social contact matrices for Irish population mixing at 5 year intervals \citep{prem2017}}
	\label{fig::cm}
\end{figure}

Contact tracing has been a prominent factor in disease suppression in Asian countries to date. However, such data is not available in an Irish context, and we are unaware of any large-scale survey or study on age-structured social mixing patterns in Ireland. However, \citet{fumanelli2012} and \citet{prem2017} provide a methodology for deriving contact patterns by leveraging  mixing patterns studied in other European countries. Our analysis relies on contact matrices given by \citet{prem2017} who projected age and location specific contact matrices in 16 age bands for 152 countries including Ireland. These are constructed from the POLYMOD study \citep{mossong2008}, which incorporated large-scale demographic household surveys (from the UN population division) and school and labour force participation rates. The estimated Irish contact interactions are shown in Fig~\ref{fig::cm}. For the purpose of our work, we sum together expected contacts in the home, work, school and other locations to give an overall matrix of expected contacts (`All' in Fig~\ref{fig::cm}). We assume that the Irish contact matrix applies to just Dublin.  

Government interventions to suppress virus spread result in changes in population mixing. Therefore, to reflect these changes, we introduce a free parameter, $\theta_k$, which scales the aggregate expected contact matrix for the time interval $\mathcal{I}_k = (r_{k},r_{k+1}]$ corresponding to the $k$th of $N$ lockdown regimes. The aforementioned age-structured contact matrices are formed by entries $c_{i\,j}$ representing the average number of daily contacts between an individual in age category $i$ with an individual in age category $j$, where $i,j = 1,\ldots,A$. Then, at time $t \in \mathcal{I}_k$ (i.e., during the $k$th lockdown), the scaled contact matrix is

$$ 
\begin{pmatrix}
	\theta_k c_{1\,1} & \ldots & \theta_k c_{1\,A}\\
	\vdots & \ldots & \vdots \\
	\theta_k c_{A\,1} & \ldots & \theta_k c_{A\,A}
\end{pmatrix} = \theta_k \mathbf{C}, \qquad k=1,\dots,N.
$$
In Section~\ref{sec::epi_model} we outline the estimation of the scaling parameters $\boldtheta  =(\theta_1,\dots,\theta_N)$ for each lockdown period using observed incidences in Dublin. This allows an estimation of macro-level behavioural changes in socialising brought about by specific measures.

\subsection{Economic cost of lockdown measures}
\label{subsec:costings}

Our age-structured modelling approach offers the potential for evaluation of age-related economic or health costs in tandem with public morbidity. Here, we explore costing economic impacts of different lockdown measures used by the Irish government to date. There has been much interest in the economic impact of COVID--19, with a rapidly expanding literature. For example, \citet{eichenbaum2020macroeconomics} study the basic macroeconomics of epidemics while \citet{brodeur2020literature} attempt a partial literature review of the main approaches to macroeconomic costing of the epidemic to date. \citet{kaplan2020great} extend the canonical macroeconomic framework to SEIR models of the type we develop in this paper. \citet{beirne2020potential} produce initial estimates of the medium-term impact of the crisis on the Irish economy. Relating to Ireland, we draw on the work of \citet{wolosko:2020} in order to construct weekly estimates of year on year (same week in different years) growth rate of Gross Domestic Product (GDP). This existing approach makes forecasts based on a neural network modelling framework, with training of the model done via quarterly GDP and Google Trends search intensity data gathered over forty six countries from the beginning of 2005 for sixty one quarters. The quarterly data is used to construct a forecasting model corresponding to a weekly resolution making an assumption of frequency neutrality~\citep{wolosko:2020}.

\par
While there are obvious criticisms of using GDP as a sound economic measure, here we employ it as a proxy to give a high level view of implied costs of lockdown to the economy. In this frame of reference, we can justify the approximation in three ways.  Firstly, GDP is an aggregate flow measure containing values in euros of expenditures by households, firms, and government, particularly the importing, and exporting parts of the Irish economy \citep{coyle2015gdp}. 
The weekly GDP tracker combines the sometimes countervailing microeconomic elements of the COVID crisis into a single variable. While expenditures by the household and corporate sectors was reduced due to government restrictions, the government expanded its support activities in terms of direct payments to furloughed workers, turnover replacements and generous liquidity packages for firms, debt and tax arrears-warehousing, and increases in direct spending on health-related measures \citep{devereux2020discretionary}.  
Secondly, it is acknowledged that Ireland is a small open economy with a very large multinational sector relative to other countries of similar size and development level. While GDP is understood for example by \citet{fitzgerald2020national} as being a relatively poor measure of the health of the Irish economy due to the influence of large multinational companies on the economy, these companies, particularly in the pharmaceutical and ICT sectors, helped keep the government's budget deficit lower than it otherwise would have been. 
Finally, the weekly GDP calculations of \citet{wolosko:2020} enable the production of a measure of the overall economic impact of COVID--19 relative to each lockdown period. A weekly tracker of economic activity allows for matching of economic predictions with the timescale of typical lockdown duration. The generation of in-model behavioural elasticities using a panel of 46 countries across several decades also means the model has the ability to react within a period of crisis, as the economy's participants learn from the first lockdown experience, and adapt themselves to cope with further lock downs. 
\section{SEIR model specification}
\label{sec::epi_model}

Irish population modelling of COVID--19 during the crisis has been carried out by the Irish Epidemiological Modelling and Advisory Group (IEMAG)~\citep{iemag2020}. They present a model for the Irish population where, at any point in time, an individual is assumed to be in one of a number of distinct model compartments that describe COVID--19 status. Movement between compartments over time is based on the current understanding of the epidemiology of COVID--19, as evidenced by the extensive literature review and evidence synthesis conducted by \citet{griffin2020rapid, mcaloon2020, byrne2020}. 
\par
We evolve this model to consider age-structured differences in population mixing. We assume closed age classes, such that population $N_i$ of age class $i$ is the sum of susceptible ($S_i$), exposed ($E_i$), infected and removed ($R_i$) compartments for that age class. There is no movement between age classes. Infected cases fall into a number of compartments: asymptomatic ($\isuper{AS}_i$), pre-symptomatic ($\isuper{PS}_i$), symptomatic and self-isolating without testing ($\isuper{SI}_i$), symptomatic and awaiting test results ($\isuper{ST}_i$), symptomatic and isolating after receiving positive test results ($\isuper{PI}_i$) and symptomatic but not tested or isolating ($\isuper{SN}_i$). The closed age class assumption implies that
\[
N_i = S_i + E_i + \isuper{AS}_{i}   + \isuper{PS}_{i} + \isuper{SI}_i + \isuper{ST}_i + \isuper{PI}_i + \isuper{SN}_i + R_i, \qquad i=1,\dots,A.\]
Exposed individuals are those incubating the disease but not yet infectious. Asymptomatic individuals are infectious but do not exhibit symptoms. Pre-symptomatic individuals are infectious but have yet to show symptoms. As pre-symptomatic individuals' symptoms develop, they will move to one of the infectious or symptomatic compartments, either self isolating and following government guidance around testing, or neither getting tested nor isolating when symptomatic, i.e., ignoring symptoms. Following infection, individuals move to the removed class ($R_{i}$), which accounts for cases who recover and those who die. 

\tikzset{%
	block/.style    = {draw, thick, rectangle, minimum height = 3em,
		minimum width = 3em},
	sum/.style      = {draw, circle, node distance = 2cm, minimum width=13mm}, 
	input/.style    = {coordinate}, 
	output/.style   = {coordinate} 
}

\begin{figure}
	\begin{center}
		\begin{tikzpicture}[scale=0.9, transform shape, auto, thick, node distance=2cm, >=triangle 45]
			\draw
			node at (0,0)[left=-30mm]{}
			node [sum, color=blue] (si) {$S_i$}
			node [sum, right=15mm, right of =si, color=blue] (ei) {$E_i$}
			node [sum, right=60mm, below = 20mm, below right of = ei, color=blue ] (iia) {$\isuper{AS}_i$}
			node [sum, right=20mm, above right of= ei, color=blue] (iip) {$\isuper{PS}_i$}
			node [sum, right =15mm, right of = iip, color=blue] (iik) {$\isuper{SI}_i$}
			node [sum, below of = iik, color=blue] (iin) {$\isuper{SN}_i$}
			node [sum, above of = iik, color=blue] (iit1) {$\isuper{ST}_i$}
			node [sum, right = 10mm,right of= iit1, color=blue] (iit2) {$\isuper{PI}_i$}
			node [sum, right=55mm, below of=iik, color=blue] (r) {$R_i$}
			node [left=2mm, below=15mm, above of=r] (ru){}
			;
			\draw[->, color=darkgray] (si) -- node {$\beta \frac{g_i(t,\bZ,\mbox{\small \boldmath{$\theta$}})}{N_i}$} (ei);
			\draw[->, color=darkgray, rounded corners=.1cm, anchor = center, midway, above right ] (ei)  |- node {$\qquad \qquad \qquad \fsub{AS}/\tausub{L}$} (iia);
			\draw[->, color=darkgray,  rounded corners=.1cm, anchor=center, midway, above right] (ei) |- node {$(1-\fsub{AS})/\tausub{L}$} (iip);
			\draw[->, color=darkgray, rounded corners=.1cm,anchor=center, midway, above right] (iip) |- node {$\qquad \frac{1-\fsub{SI}-\fsub{T}}{\tausub{C}-\tausub{L}}$} (iin);
			\draw[->, color=darkgray] (iip) -- node {$\frac{\fsub{ST}}{\tausub{C}-\tausub{L}}$} (iik);
			\draw[->, color=darkgray,  rounded corners=.1cm,anchor=center, midway, above right] (iip) |- node {$\,\,\qquad \frac{\fsub{T}}{\tausub{C}-\tausub{L}}$} (iit1);
			\draw[->, color=darkgray] (iit1) -- node {$\frac{1}{\tausub{R}}$} (iit2);
			\draw[->, color=darkgray,  rounded corners=.1cm,anchor=center, midway, above left] (iia) -| node {$1/\tausub{D}\qquad \qquad \quad$} (r);
			\draw[->, color=darkgray] (iin) -- node {$\frac{1}{\tausub{D} - \tausub{C} + \tausub{L}}\,$} (r);
			\draw[->, color=darkgray, rounded corners=.1cm,anchor=left, above left = 20mm] (iik) -| node {$\frac{1}{\tausub{D} - \tausub{C} + \tausub{L}}\qquad \qquad  \,\,$} (ru);
			\draw[->, color=darkgray,rounded corners=.1cm,anchor=center, midway, above] (iit2) -| node {$\frac{1}{\tausub{D} - \tausub{C} + \tausub{L} - \tausub{R}}\qquad\,\,$} (r);
			
		\end{tikzpicture}
	\end{center}
	\caption{Diagram representing interactions in the age-structured SEIR system of ODEs with rate of movement between classes indicated. A full description is given in Equation~(\ref{eq::ode}).}
\end{figure}
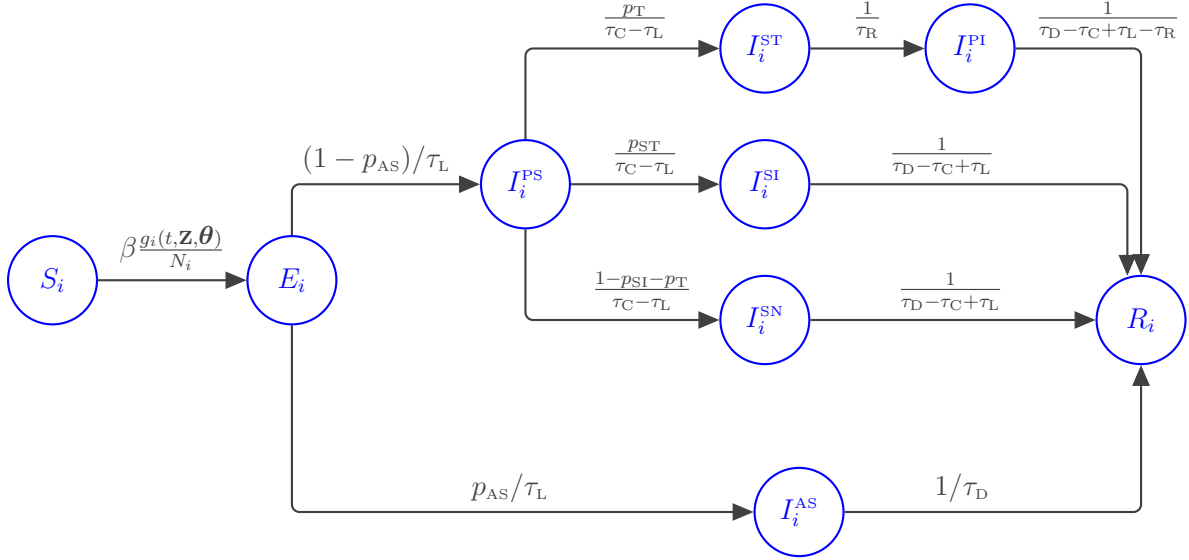

We write the system of ordinary differential equations (ODEs) describing the SEIR model for age class $i=1,\dots,A$:
\begin{equation}
	\begin{split}
		& \begin{split}
			& \deriv{{S}_i}{t}  = -\beta\, \frac{g_i(t,\mathbf{z},\boldtheta)}{N_i}  \,S_i\\
			& \\
			&\deriv{\isuper{AS}_i}{t}  = \fsub{AS} \,\frac{E_i}{\tausub{L}} \,-\,  \frac{\isuper{AS}_i}{\tausub{D}}\\
			& \\
			&\deriv{\isuper{SI}_i}{t}  = \fsub{SI}\,\frac{\isuper{PS}_i}{\tausub{C} - \tausub{L}} \, -\, \frac{ \isuper{SI}_i}{\tausub{D} - \tausub{C} + \tausub{L}} \\
			& \\
			& \deriv{\isuper{SN}_i}{t} = (1 - \fsub{SI} - \fsub{T})\,\frac{\isuper{PS}_i}{\tausub{C} - \tausub{L}} \, -\, \frac{\isuper{SN}_i}{\tausub{D} - \tausub{C} + \tausub{L}}  
		\end{split}
		\qquad \qquad
		\begin{split}
			& \deriv{{E_i}}{t}  = \beta\, \frac{g_i(t,\mathbf{z},\boldtheta)}{N_i}\,S_i - \frac{E_i}{\tau_{\mbox{\tiny L}}}\\
			& \\
			& \deriv{ \isuper{PS}_i }{t} = (1-\fsub{AS})\, \frac{E_i}{\tausub{L}} \,-\, \frac{\isuper{PS}_i}{\tausub{C} - \tausub{L}} \\
			&\\
			& \deriv{\isuper{ST}_i}{t}= \fsub{T}\,\frac{\isuper{PS}_{i}}{\tausub{C} - \tausub{L}}\,  -\, \frac{\isuper{ST}_{i}}{\tausub{R}} \\
			& \\
			& \deriv{\isuper{PI}_i}{t} = \frac{\isuper{ST}_{i}}{\tausub{R}}  \,-\, \frac{\isuper{PI}_{i}}{\tausub{D} - \tausub{C} + \tausub{L} - \tausub{R}} 
		\end{split}\\
		& \\
		& \qquad \qquad \deriv{R_i}{t} = \frac{\isuper{AS}_i}{\tausub{D}} \, +\, \frac{\isuper{SI}_i}{\tausub{D} - \tausub{C} + \tausub{L}}  \,+ \,\frac{\isuper{PI}_i}{\tausub{D} - \tausub{C} + \tausub{L} - \tausub{R}}  \,+\, \frac{\isuper{SN}_i}{\tausub{D} - \tausub{C} + \tausub{L}}   
	\end{split} \label{eq::ode}
\end{equation}
where the function $g_i(t,\mathbf{z}, \boldtheta)$ in the mass relation for being exposed when susceptible is 
\[
g_i(t,\mathbf{z},\boldtheta) = \sum_{k=1}^N \,\mathbb{I}\left( \, t \in \mathcal{I}_k\,\right)\,\sum_{j=1}^A \, \theta_k \,c_{ij} \, \left[ \, \alpha \,\isuper{AS}_j + \isuper{PS}_j + \kappa\,  \isuper{SI}_j + \isuper{ST}_j+  \kappa \, \isuper{PI}_j + \isuper{SN}_j\, \right]
\]
where $\mathbf{z}$ denotes the entire state vector \[\mathbf{z}=\left(\,S_1,E_1, \dots, S_2, E_2, \dots, S_A, E_A, \dots, R_A \,\right),\] and $\mathbb{I}\left( \, t \in \mathcal{I}_k\,\right)$ is an indicator function which equals one when $t \in \mathcal{I}_k$ and is zero otherwise. Susceptible individuals in age class $i$ are exposed to the virus through contacts with infected individuals in all age classes and this is described through the function $g_i(t,\mathbf{z},\boldtheta)$. The level of exposure is modulated by the scaled average number of daily contacts with each age class, with a scaling factor for each lockdown regime. The parameters $\alpha$ and  $\kappa$ are used to describe transmission dynamics with respect to the various infected compartments. The reduction of transmission due to being asymptomatic is given by $\alpha$ which is taken as $0.55$~\citep{evoy2020}. Individuals in isolation are  expected to have a reduced infectious burden on those they interact with. This is described using $\kappa = 0.05$~\citep{iemag2020} and we assume an identical reduced burden from those who are positive and isolating. 
\par
The SEIR model makes extensive use of $\beta$, the multiplicative force of infection. $\beta$ is chosen based on a specified value of $R_0$, the baseline reproduction rate. Baseline here highlights that this is the expected reproduction rate in Dublin, with a fully susceptible population (i.e., $S_i = N_i$, $i=1,\dots,A$), under no intervention. $R_0$ can be expressed as the largest absolute eigenvalue of the next generation matrix, $\mathbf{Q}= \mathbf{F}\,\mathbf{V}^{-1}$ \citep{diekmann1990}, where $\mathbf{F}$ and $\mathbf{V}$ are block matrices which describe the transmissions and the transitions between compartments, respectively. The analytic form of these matrices is given in supplementary material B. Factoring $\beta$ out of the $\mathbf{F}$ matrix, $\mathbf{F}= \beta\, \widehat{\mathbf{F}}$,  the dominant eigenvalue of $\mathbf{Q}$ can be expressed as a product of $\beta$ and the dominant eigenvalue of $\widehat{\mathbf{F}}\,\mathbf{V}^{-1}$. Hence, if $R_0$ is determined and $\beta$ is desired, the expression can easily be rearranged to give $\beta$ as follows: 
\begin{equation}
	\beta = R_0/\xi
\end{equation}
where $\xi$ is the largest eigenvalue of $\widehat{\mathbf{F}}\, \mathbf{V}^{-1}$. Note, the values of $R_0$ and $\beta$ are only calculated once as they are baseline values; shifts in infection dynamics away from the baseline are captured by $\boldtheta$. Our choice for $R_0$ is 3.4, based on \citet{naraigh2020}. This leads to $\beta = 0.031$ when not under intervention. 
A full list of parameter value settings used in our modelling is given in supplementary material Table~\ref{tab::par}. We note here that the $\theta_k, k=1,\dots,N$ are unknown. The next section describes how these are estimated using observed case incidences.

\section{Model Fitting} \label{sec::model_fit}

Specification of the model in Section~\ref{sec::epi_model} uses parameters $\boldtheta=(\theta_1,\dots,\theta_N)$ to rescale, for each of the $N=12$ lockdown intervention policies, what would have been the assumed average contacts between individuals in the various age classes under normal circumstances (i.e., prior to the pandemic). Estimation of these parameters is of interest in predicting behaviour during lockdown, and hence for forecasting the benefit of specific interventions. Note that the remaining parameters (i.e., those other than $\boldtheta$) on which the dynamics of the SEIR model (\ref{eq::ode}) depend, describing the flow of individuals between compartments, are based on expert opinion~\citep{iemag2020,naraigh2020}. The data currently available is not sufficiently rich to estimate these. We first describe estimation of $\boldtheta$ and then a parametric bootstrap method \citep{Pawitan05} to provide uncertainty, which can be propagated through model forecasts.
\par
In order to link the model with observed data we monitor the cumulative number of cases up to time $t$ for each age class. It is only cases exiting the $\isuper{ST}_i, i=1,\dots,A$ compartment that can be linked to observed incidence counts in the general population. We can think of a variable, counting infected cases as they exit compartment $\isuper{ST}_i$ before going into $\isuper{PI}_i$. For scaling $\boldtheta$, age class $i$ and time $t$, denote this by $X_i(t;\boldtheta)$. This can be related to the other model compartments through
\begin{equation} \label{eq::count_comp}
	\deriv{X_i}{t} = \frac{\isuper{ST}_i}{\tausub{R}},\qquad i=1,\dots,A.
\end{equation}

To compare outputs from the SEIR model with observed data, we use this count aggregated over age classes:
\begin{equation} \label{eq::count_sum}
	X(t;\boldtheta) = \sum_{i=1}^A X_i(t;\boldtheta),
\end{equation}
which gives total cumulative case counts to time $t$. Evaluating this at $t_d = hd, \,d=1,\dots,n$ where $h$ generates a time discretisation corresponding to consecutive days, we can then compare $X(t_d;\boldtheta)$ to observed cumulative cases at day $d$. We denote the observed cumulative counts at day $d$ by $x_d$. 

\subsection{Estimation of regime specific contact scaling parameters} \label{subsec::par_op}
To estimate $\boldtheta$, we minimize the squared error loss, i.e., the residual sum of squares,
\[
RSS(\boldtheta) = \sum_{d=1}^n \left(x_d - X(t_d;\boldtheta)\right)^2
\] 
on cumulative case counts. Minimization is carried out using the default Nelder-Mead alogrithm~\citep{nelder1965simplex} provided in the \texttt{R} package \texttt{optimx} \citep{nash2011unifying}. Note that each step of this algorithm, corresponding to a proposed $\boldtheta$ vector, requires the calculation of $X(t;\boldtheta)$ to evaluate the suitability of $\boldtheta$ through $RSS(\boldtheta)$. In order to obtain $X(t;\boldtheta)$, the system of ODEs given in (\ref{eq::ode}) are numerically solved using the \texttt{R} package \texttt{deSolve} \citep{soetaert2010solving}. Specifically, we have found the \texttt{lsoda} function within this package to be particularly flexible, providing automatic selection of stiff or non-stiff methods; see \citet{nash2011unifying} for details. When solving the system of ODEs for a candidate $\boldtheta$, we take $\isuper{PS}_i=1/A$, $E_i=15/A$ and hence $S_i=N_i - 16/A$ as the initial values for $i = 1,\dots,A$, as per \citet{iemag2020}. The initial values for the remaining compartments are set to 0. Multiple random initialisations of $\boldtheta$ are used to improve robustness of the overall algorithm with respect to the issue of convergence to a local minima. When generating these initial vectors, we assume that the effect of lockdown measures  is to reduce social mixing below pre-pandemic levels, and, therefore,  use a $U(0,1)$ draw to initialise each parameter, i.e., $\boldtheta_{k}^{0}\sim U(0,1), k = 1,\dots,N$. A summary of our estimation procedure is given in Algorithm \ref{alg::est}. 



\begin{algorithm}
	\caption{Estimation of $\boldtheta$}\label{alg::est}
	\begin{algorithmic}[1]
		\Procedure{estimation}{$M$,\,$x_d$,\,$d=1,\dots,n$}
		\For{$m = 1,\dots, M$}
		\State $\boldtheta_k^{0(m)} \sim \mathrm{Unif}(0,1)$,\qquad$k=1,\ldots,N$
		\State $RSS(\boldtheta) := \sum_{d=1}^n ( x_d - X(t_d;\boldtheta))$ with $X(t_d;\boldtheta)$ given by \texttt{lsoda}
		\State $\widehat\boldtheta^{(m)}= \operatorname*{argmin}_\theta RSS(\boldtheta)$ 
		using \texttt{optimx} initialised at $\boldtheta^{0(m)}$  
		\State $RSS^{(m)} = RSS(\widehat\boldtheta^{(m)})$
		\EndFor
		\Return{$\widehat{\boldtheta} = \{\widehat{\boldtheta}^{(m)} \mid RSS^{(m)} = \min(RSS^{(1)},\ldots,RSS^{(M)})\}$}  
		\EndProcedure
	\end{algorithmic}
\end{algorithm}
\par
Since we have to solve the ODEs (\ref{eq::ode}) numerically using \texttt{lsoda} of each iteration within the \texttt{optimx} optimisation, the above procedure is computationally intensive. On average it takes approximately one hour to run the optimisation for each random initialisation, and we have used $C = 300$ random initialisations. Thus, to improve the computational feasibility, we have run initialisations in parallel on an EC2 instance hosted by Amazon Web Services with $32$ cores and $64$ GB memory.  

\subsection{Propagating uncertainty in contact scaling parameters}\label{subsec::uncert}

We explore uncertainty in the estimation of $\boldtheta$ and investigate how this propagates into the reproductive rate. In order to quantify uncertainty we follow~\citet{chowell2017} by using a parametric bootstrap approach. This approach makes use of an assumed generative parametric model for daily case counts based on the observed case counts. This model is then used to re-generate $B$ synthetic instances of the daily new case series; each of these instances is used to re-estimate the vector $\boldtheta$. The resulting empirical distribution of the re-estimated vectors can be used as an approximation to the sampling distribution of $\widehat{\boldtheta}$ (the estimate based on the original cumulative case counts).
\par
The estimate $\widehat{\boldtheta}$ is found using the observed daily cumulative counts as described in Section~\ref{subsec::par_op}. Given this estimate, the expected daily case count $\widehat{\mu}_d$ for day $d$ can be predicted using
\[
\widehat{\mu}_d = X(t_{d};\widehat{\boldtheta}) - X(t_{d-1};\widehat{\boldtheta}), \qquad d \ge 1
\]
where $t_0 := 0$ and $X(0;\boldtheta) := 0$. We assume a negative binomial distribution~\citep{chowell2017} as a generative model for daily case counts $Y_d \sim \mathrm{NegBin}(\widehat{\mu}_d,\rho)$ with expected value $\widehat{\mu}_d$ and dispersion parameter $\rho$:

\begin{eqnarray}	\label{eq:negbin}
	\mathrm{Pr}\left(Y_d = y\right) &=& \frac{\Gamma(\rho+y)}{y!\,\Gamma(\rho)}\left(\frac{\widehat{\mu}_d}{\widehat{\mu}_d+\rho}\right)^{y}\left(1+\frac{\widehat{\mu}_d}{\rho}\right)^{-\rho}
\end{eqnarray}
where we have parameterised the negative binomial distribution through its expected value and dispersion. The value of $\rho$ used for generating bootstrap $Y_d$ series is the maximum likelihood estimate $\hat{\rho}$ based on the observed daily cases.
\par
For each of $b=1,\dots,B$ bootstrap replications, we generate daily counts $y_d^{(b)}$ and convert to cumulative counts $x_d^{(b)},\,d=1,\dots,n$. Then the method of Section~\ref{subsec::par_op} is applied to the $x_d^{(b)}$ to produce a bootstrap estimate $\widehat{\boldtheta}^{(b)}$. Collectively, the $B$ estimates $\widehat{\boldtheta}^{(b)}$ provide an approximation to the sampling distribution of $\widehat{\boldtheta}$. The steps are summarized in Algorithm~\ref{alg::boot}.

\begin{algorithm}
	\caption{Parametric bootstrapping}\label{alg::boot}
	\begin{algorithmic}
		\Procedure{bootstrap}{$B$,\,$\widehat{\rho}$,\,$\widehat{\mu}_d, d=1,\dots,n$}
		\For{$b = 1,\dots, B$}
		\State $x_0^{(b)} := 0$
		\For{$d=1,\dots, n$}
		\State $Y_d \sim \mathrm{NegBin}(\widehat{\mu}_d,\widehat{\rho})$
		\State $x_d^{(b)} = x_{d-1}^{(b)}+ y_d$
		\EndFor
		\State $\widehat{\boldtheta}^{(b)}$ obtained from Algorithm \ref{alg::est}
		\EndFor
		\Return{Bootstrap sample $\{\widehat{\boldtheta}^{(b)}, 1, \ldots, B\}$}
		\EndProcedure
	\end{algorithmic}
\end{algorithm}

\section{Results} \label{sec::results}
In this section we present the results of fitting the age-structured SEIR model to the Irish data using the methodology described in Section \ref{sec::model_fit}, and also discuss some economic findings.

\subsection{Fitted SEIR model} \label{sec:fittedSEIR}

\begin{figure}[ht]
	\centering
	\includegraphics[scale = 0.52]{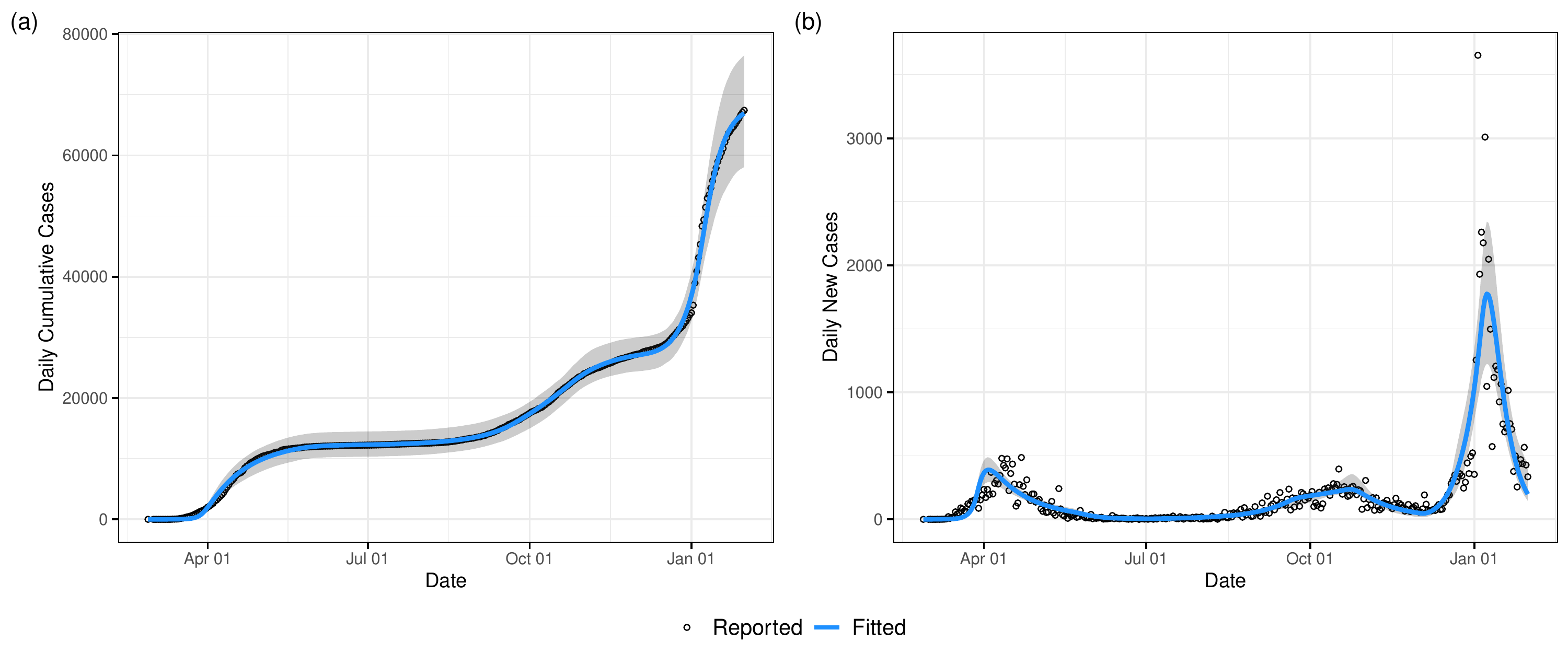}
	\caption{Model fit to daily recorded cases with bootstrapped 95\% uncertainty bounds.}
	\label{fig::model_fit}
\end{figure}

Fig~\ref{fig::model_fit}(a) shows the model fit to the daily cumulative cases data (i.e., $X(t_d;\widehat\boldtheta)$ and $x_d$ respectively), while Fig~ \ref{fig::model_fit}(b) shows a plot of the model fit to the daily new cases data (i.e., $X(t_d;\widehat\boldtheta) -  X(t_{d - 1};\widehat\boldtheta)$ and $x_d - x_{d-1}$ respectively). Both figures illustrate that the model provides a good fit to the observed data albeit with slight deviations in the early days of the epidemic and later around the December holiday and New Year period. We observe from the model fit to daily new cases that these periods were characterised by larger variability in the daily recorded number of new cases. The presence of outliers may be as a result of data reporting issues, especially over the December holiday period. For example, in Ireland a testing backlog developed over this period with test results from multiple days being subsequently batched together.


\par
\begin{figure}[h]
	\centering
	\includegraphics[scale= 0.72]{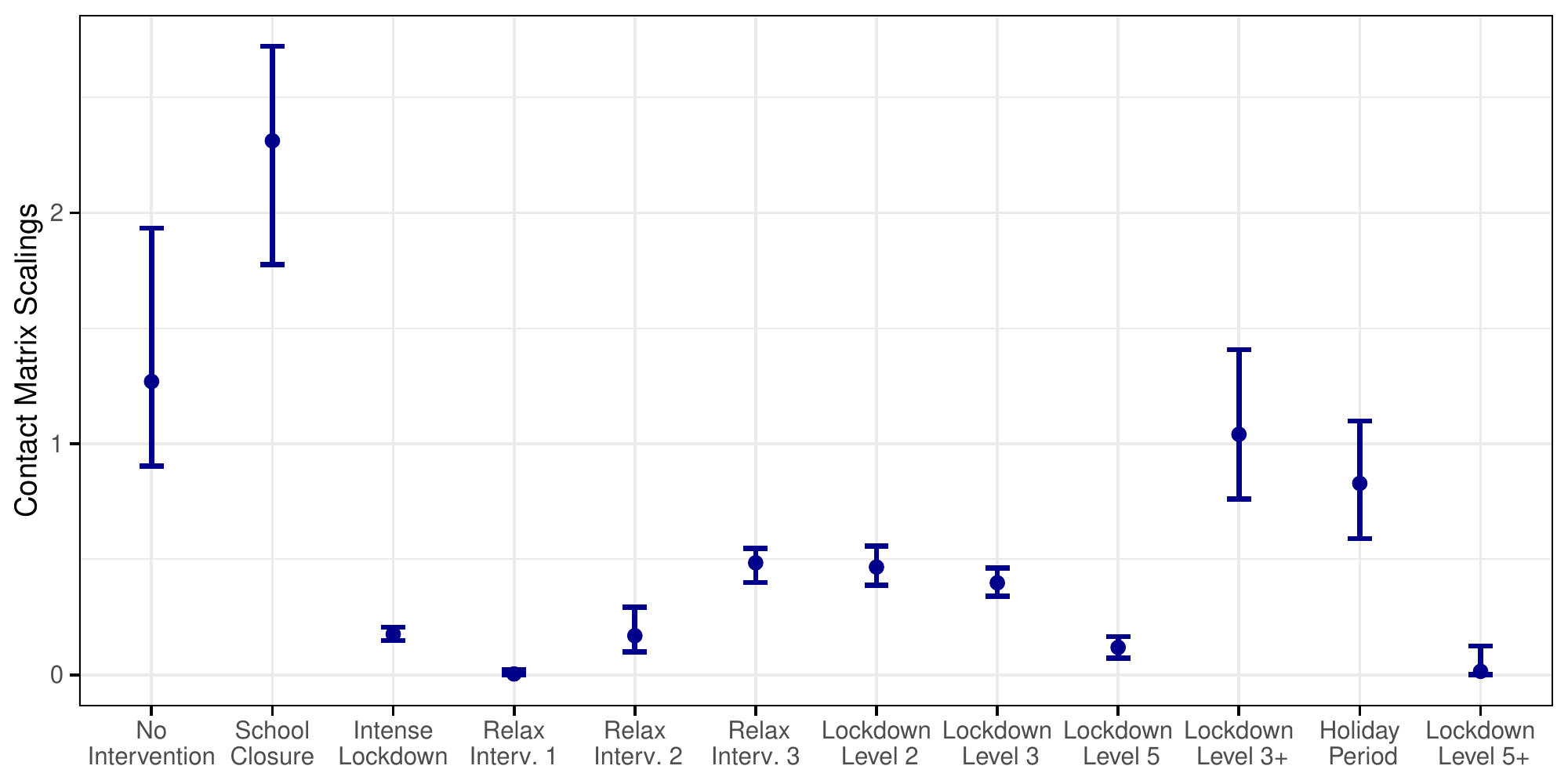}
	\caption{Estimates of the scaling parameters with bootstrapped 95\% uncertainty bounds.}
	\label{fig::theta_err_bars}
\end{figure}

Fig~\ref{fig::theta_err_bars} displays the estimated scaling parameters $\boldtheta$ with bootstrapped $2.5\%$  and $97.5\%$ percentiles for each government policy observed over the period of study. The associated numeric values are deferred to Table S2 supplementary material. 
\par
The scaling parameter estimate for the no-intervention period, $\widehat\theta_1$ $\approx$ 1.27  ($95\%$ CI 0.90 - 1.94), indicates that the social contact patterns based on the POLYMOD study may be slightly under-estimating the current Irish contact patterns. Perhaps somewhat surprisingly, the parameter $\widehat\theta_2$, corresponding to an initial school closure period, is greater than one ($95\%$ CI 1.78 - 2.72). However, this might be explained by the fact that it corresponds to a short period of time where no other measures had yet been introduced (apart from pub closures later in the period), but with an imminent government announcement of a strict nationwide lockdown expected -- in line with what had been observed in other countries already by this stage in the global pandemic. During this period there was frenzied panic buying and stock piling of goods, increased travel across the country, and possibly increased social gatherings prior to movement restrictions. 
The remaining scaling parameters behave as expected based on the level of restrictions in place at that time: the higher the levels of restrictions, the smaller the scaling parameter, corresponding to reduced social mixing. The confidence intervals just prior to and including the holiday period indicate that social mixing returned to a near normal level at that time where a dramatic spike in the case numbers was also observed. This was followed immediately by a heavy lockdown and consequent drop in case numbers; indeed, the scaling parameter for this final period has the smallest value of all. 
\par
Over the period of study, note that we have witnessed two lockdown Level 3 periods (September/October and December 2020). However, although in theory both periods were designated as ``lockdown Level 3'' by the Irish government, we have applied two separate scaling parameters for these two periods as the December lockdown included some relaxations compared to a full Level 3 lockdown. Specifically, non-essential retail and services were open once again and indoor service in restaurants and cafes was also permitted. This was done to facilitate people's social needs around the holiday period, and indeed we see that the estimated parameter value for lockdown Level 3+ (December) is much larger than that for lockdown Level 3 (September/October). Again because of modifications in the execution, we have separate scaling parameters for the lockdown Level 5 in October/November and what we call the lockdown Level 5+ January 2021; the latter was stricter following the large rise in case numbers during the holiday period, and this is reflected in the small scaling value for this period as previously mentioned. Since the lockdown levels commenced with (what we label as) a Level 2 in August 2020, we have not observed a Level 1 or 4 lockdown. Prior to August 2020, the lockdowns and relaxations were more ad-hoc and do not fit into any particular governmental lockdown level. The contact matrices corresponding to the policy interventions based on our fitted model are displayed in Fig~\ref{fig::scaled_cm} in the supplementary material. 


\par
\begin{figure}[h]
	\centering
	\includegraphics[scale=0.6]{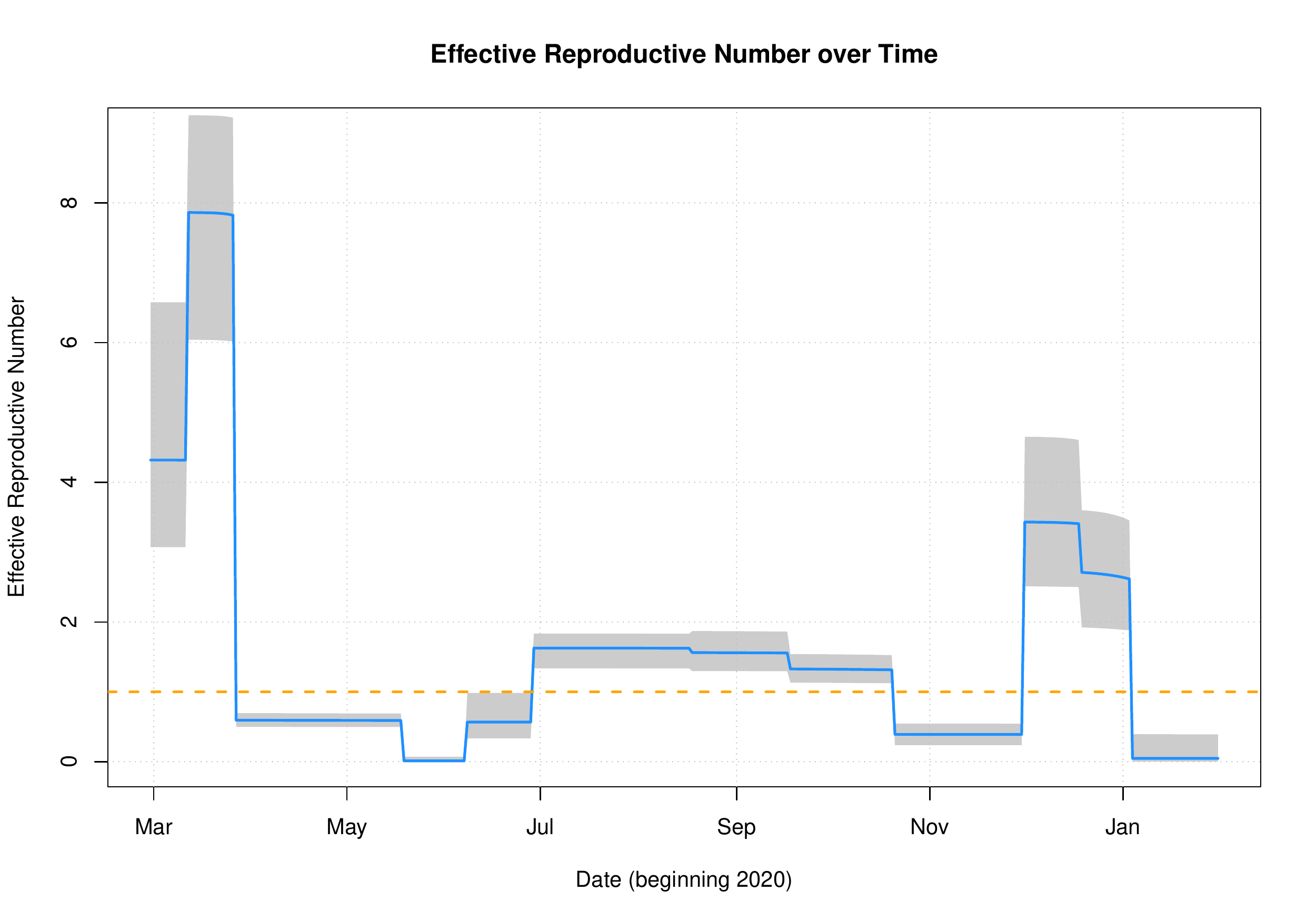}
	\caption{The effective reproduction number $R(t)$. The solid line was calculated from the best fit parameters, and the uncertainty intervals were drawn by computing $R(t)$ for each bootstrap replicate and selecting the 97.5th and 2.5th percentile at each time point.}
	\label{fig::Rt}
\end{figure}

An important epidemic metric is the \emph{effective reproductive number}, $R(t)$, which describes the expected number of secondary infections at time $t$. It differs from $R_0$, the \emph{baseline} reproduction number, in that it changes over time and takes into account that the whole population will not be fully susceptible. $R(t)$ can be calculated as the product of $R_0$ and the total proportion of susceptibles in the population at time $t$ (see for example \citet{nishiura2009}). In the context of our model, the baseline reproductive number is $R_0 \theta_k$ (rather than just $R_0$) to account for the rate at which individuals interact with each other; recall from Section \ref{sec:agesocial} that $\theta_k$ is the scaling parameter corresponding to the time interval $\mathcal{I}_k = (r_{k},r_{k+1}]$. Therefore, the effective reproductive number is
\begin{equation} \label{eq::Rt_eq}
	R(t) = R_0\theta_k \widetilde{S}(t),
\end{equation}
where $\widetilde S(t) = \sum_{i=1}^AS_i(t) / \sum_{i=1}^AN_i(t)$ is the proportion of susceptible individuals in the the entire population at time $t$. Fig~\ref{fig::Rt} displays the estimate of $R(t)$ based on our fitted model. An $R(t) > 1$ implies the infections will continue to grow exponentially, whereas $R(t) < 1$ implies infections are declining to zero; $R(t) \approx 1$ indicates a constant infection level. We see here that prior to the first heavy lockdown in April, $R(t)$ was initially very large. This dropped below one following that first lockdown, but gradually increased again over the summer period when relaxations were introduced. It was brought back under control with successive Level 3 and Level 5 lockdowns, but markedly increased over the run-up to the December holiday period; $R(t)$ was then driven towards zero with the lockdown Level 5+.\\

\subsection{Assessing the economic cost of policy interventions}
\label{sec:costings}

The approach of \citet{wolosko:2020} provides us with a trained model to estimate weekly GDP growth during the pandemic. The inputs to the model are the $\Delta s_{l,w}$, year on year differences in weekly log search intensities on Google Trends for $l=1,\dots,24$ search terms over $w=1,\dots,52$ weeks. The model has a country specific fixed effect resulting in country targeted forecasts (denoted here by ``ire'' for Ireland). The inputs are passed to a nonlinear function $\widehat{f}(\cdot,\cdot)$, which has been trained using historic data as described in Section~\ref{subsec:costings}. This results in 
\[
\widehat{g}_{\mathrm{ire},w} = \widehat{f}\left( \left\{ \Delta s_{l,w}\right\}_{l,w}, \mbox{``ire''} \right)
\]
where $\widehat{g}_{\mathrm{ire},w}$, is a forecast of GDP growth for week $w$ in Ireland.
\par
\begin{figure}[!htb]
	\centering
	\includegraphics[width=5.5in]{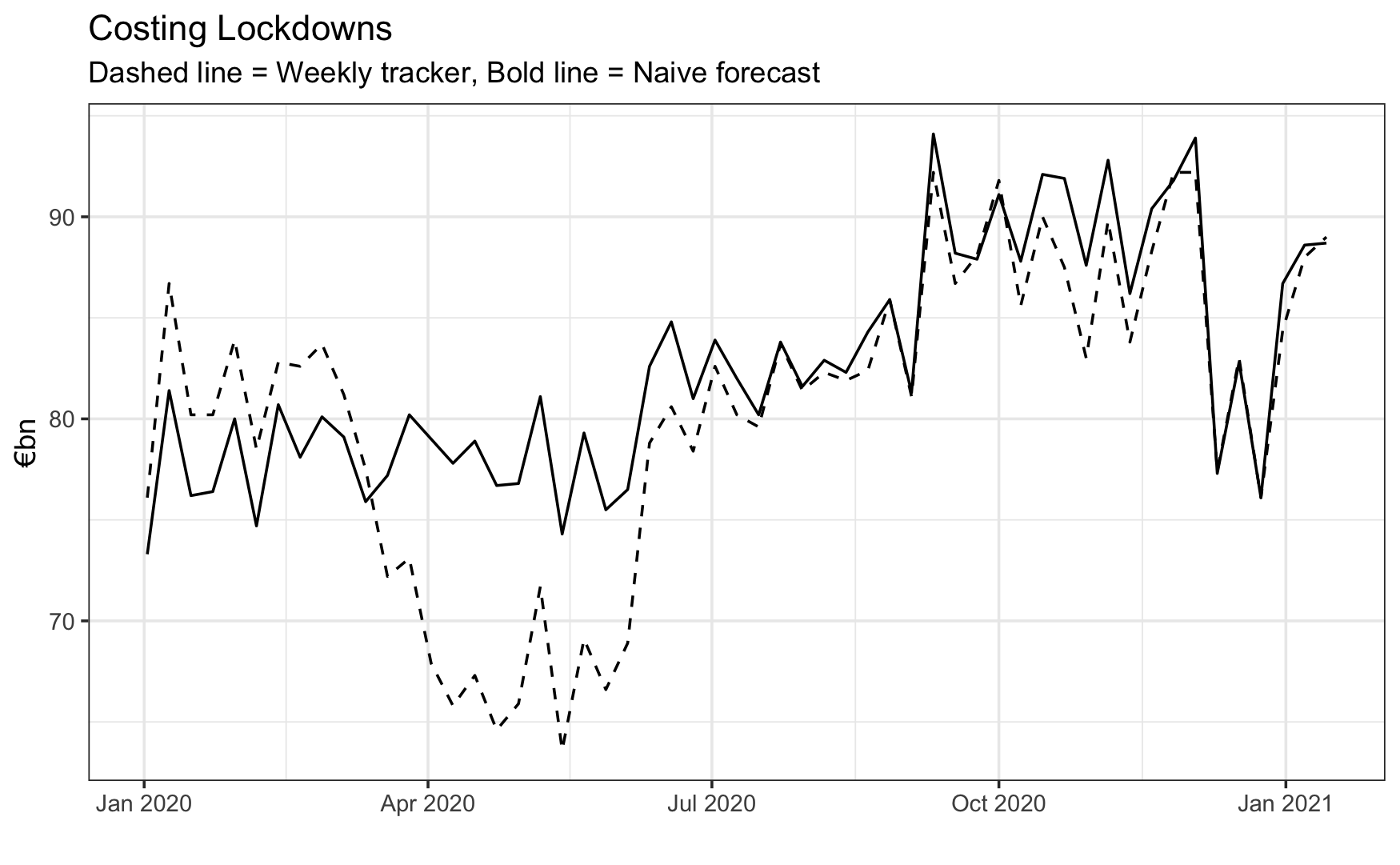}
	\caption{Comparison of weekly GDP estimates in Ireland for the naive no-pandemic ARIMA(1,1,1) (bold) and weekly tracker (dashed) approaches.}
	\label{fig:costs}
\end{figure}

Assuming weights equal to 1/48 (due to a 48 week working year) and an average counterfactual annualised growth of $5.8\%$ given by a naive forecast, the impact on growth rates per week can be calculated as
$\frac{1}{48}(5.8-\widehat{g}_{\mathrm{ire},w})$. Multiplying these by 2019 measures of GDP in euros gives an estimated cost per week in nominal euros.  A counter-factual ``no  COVID'' scenario can be constructed by naively forecasting the 2019 estimated weekly growth rate forward using an ARIMA(1, 1, 1) model to provide $\tilde{g}_{\mathrm{ire},w}$, with GDP in nominal euros is estimated in the same way. Fig~\ref{fig:costs} shows the weekly estimates of GDP for each approach, with the difference in weekly GDP figures between the approaches suggesting an estimate of the economic cost per week of pandemic related measures. Linking estimated weekly costs to the lockdown strategies implemented results in the estimated economic costs shown in Table~\ref{tab:costs}.

\begin{table}[!htb]
	\centering
	\begin{tabular}{lll|r}
		\hline
		\textbf{Start} & \textbf{End} & \textbf{Policy}         & Cost (\euro bn) \\
		\hline
		29/02/2020     & 11/03/2020   & No Intervention         & -2.47    \\
		12/03/2020     & 26/03/2020   & School Closure          & -3.68    \\
		27/03/2020     & 18/05/2020   & Intense Lockdown        & 37.14    \\
		19/05/2020     & 07/06/2020   & Relax Intervention 1    & -11.74   \\
		08/06/2020     & 28/06/2020   & Relax Intervention 2    & 5.07     \\
		29/06/2020     & 17/08/2020   & Relax Intervention 3    & 4.77     \\
		18/08/2020     & 17/09/2020   & Lockdown Level 2        & 3.69     \\
		18/09/2020     & 20/10/2020   & Lockdown Level 3        & 0.90     \\
		21/10/2020     & 30/11/2020   & Lockdown Level 5        & 8.03     \\
		01/12/2020     & 18/12/2020   & Lockdown Level 3+       & 1.30     \\
		19/12/2020     & 03/01/2021   & Holiday period          & 1.67     \\
		04/01/2021     & 31/01/2021   & Lockdown Level 5+       & 5.00     \\ 
		\hline
	\end{tabular}
	\caption{Estimated economic costs for each lockdown period.}
	\label{tab:costs}
\end{table}

\par
This simple analysis shows that relative to the counterfactual ``no COVID'' experiment, the majority of cost to society was incurred during the first intense lockdown period. Recoveries tended to overshoot the naive forecasted values (hence the negative readings in Table \ref{tab:costs} for the relaxation of interventions) for a number of reasons. For example, pent up demand in the household sector, once released, is very easily produced and consumed. The economy was also supported by large-scale government intervention which began at scale in May and June of 2020 involving government injection of cash directly into the economy to support households and firms.
Participants in the economy adapted and adjusted to their changed circumstances. This is why the calculation of elasticities in \citet{wolosko:2020} is important to include. Once people moved their consumption online, and government supports were in place, the impact of the crisis was substantially attenuated. 

\section{Forward projection and counterfactual scenario analysis\label{sec:fwdcounter}}

A key strength of our modelling framework is that it enables forward projection, and quantification of projection uncertainty, of anticipated case numbers under specific health interventions. In the following we simulate epidemiological and economic costs of a number of counterfactual scenarios based on hypothetical policy decisions taken by the Irish Government. We compare these to observed epidemiological data for that period to evaluate predictive performance. Table~\ref{tab::HPSC_cases_and_deaths} in the supplementary material presents the Irish Health Protection and Surveillance Centre's (HPSC's) total number of cases and deaths up to $28^{\text{th}}$ December 2020 \citep{HPSC2020}. The data was no longer broken down by age category to the same granularity after that date. From this we can establish approximate estimates of death rates for each age class that we can apply to the forecasts of the SEIR model.


\subsection{Projected cases, economic costs and deaths}\label{sec::resultsprojections}

Evaluation of the health impact of government restrictions on population mobility requires a contact scaling matrix for lockdown levels 1 to 5. However, as described in Section \ref{sec:fittedSEIR}, Level 1 and Level 4 lockdowns have not been used to date (and hence not observed), while Level 3 and Level 5 lockdowns were used twice but with modifications on each of the second occasions. Moreover, seasonal shifts in social dynamics (for example, in December), may play a role in the effectiveness of a given lockdown. Our chosen scaling estimates for projection, based on supplementary Table \ref{tab::est_lockdown_scalings}, make a pragmatic best approximation. For Level 0 (No Intervention), we use the `No intervention' scaling from the fitted model; Level 1 (unobserved) uses a linear interpolation of scalings from Level 0 and 2; Level 3 was enacted twice with slight changes the second time, but we use the scaling from the October 2020 lockdown (`Lockdown Level 3'); Level 4 was unobserved, so we use again a linear interpolation, this time from Level 3 to Level 5; Level 5 was used twice, but similar to Level 3 case, we use the scaling from November 2020 corresponding to `Lockdown Level 5'. The resulting scaling parameter estimates mapped to lockdown levels are presented in Fig~\ref{fig::cm_scales_2} in the supplementary material for completeness, and for reference we list values in supplementary Table~\ref{tab::interpolated_scalings}.
\par
\begin{figure}[ht]
	\begin{subfigure}
		\centering
		\includegraphics[scale=0.45]{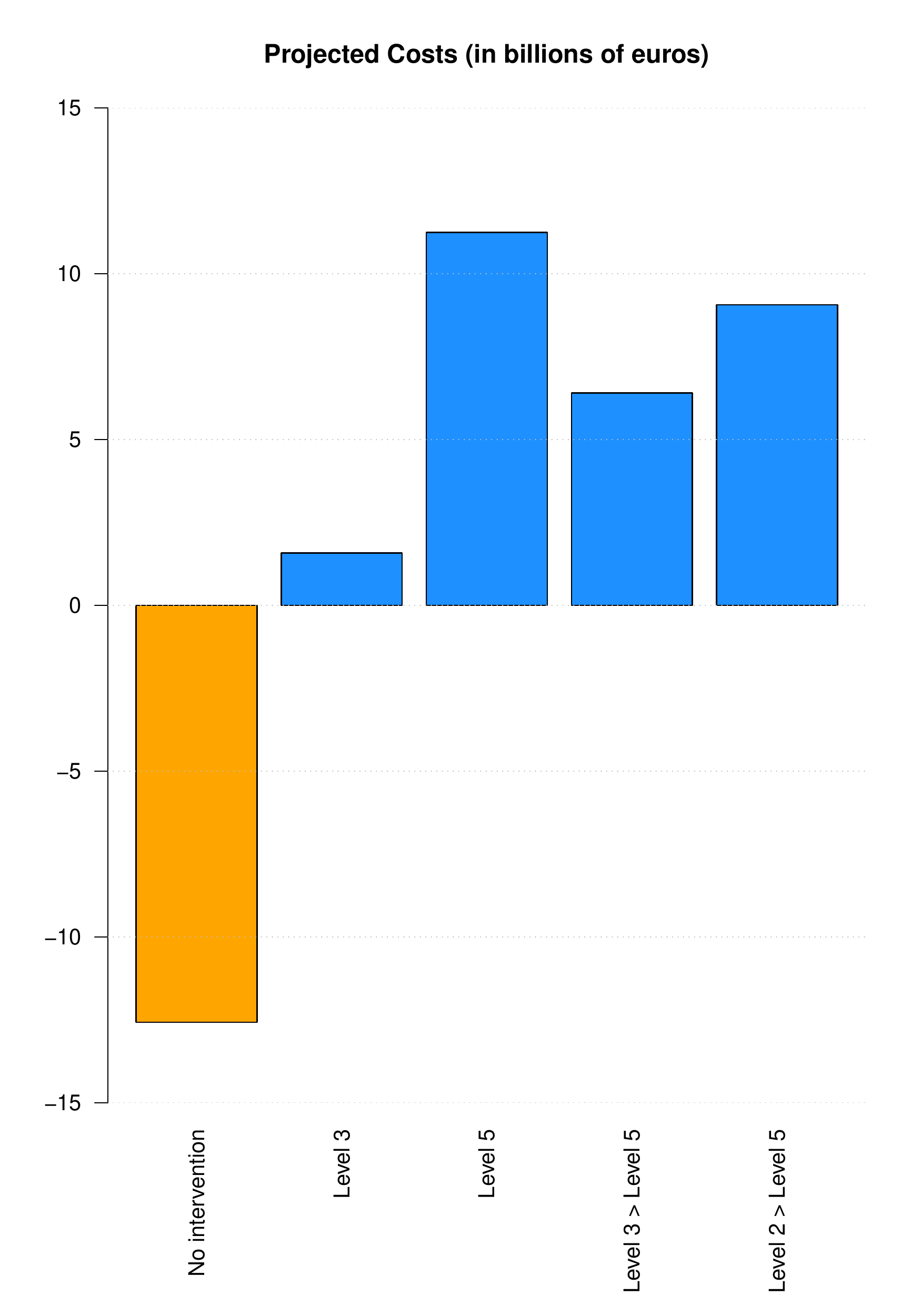}
	\end{subfigure}
	\begin{subfigure}
		\centering
		\includegraphics[scale=0.46]{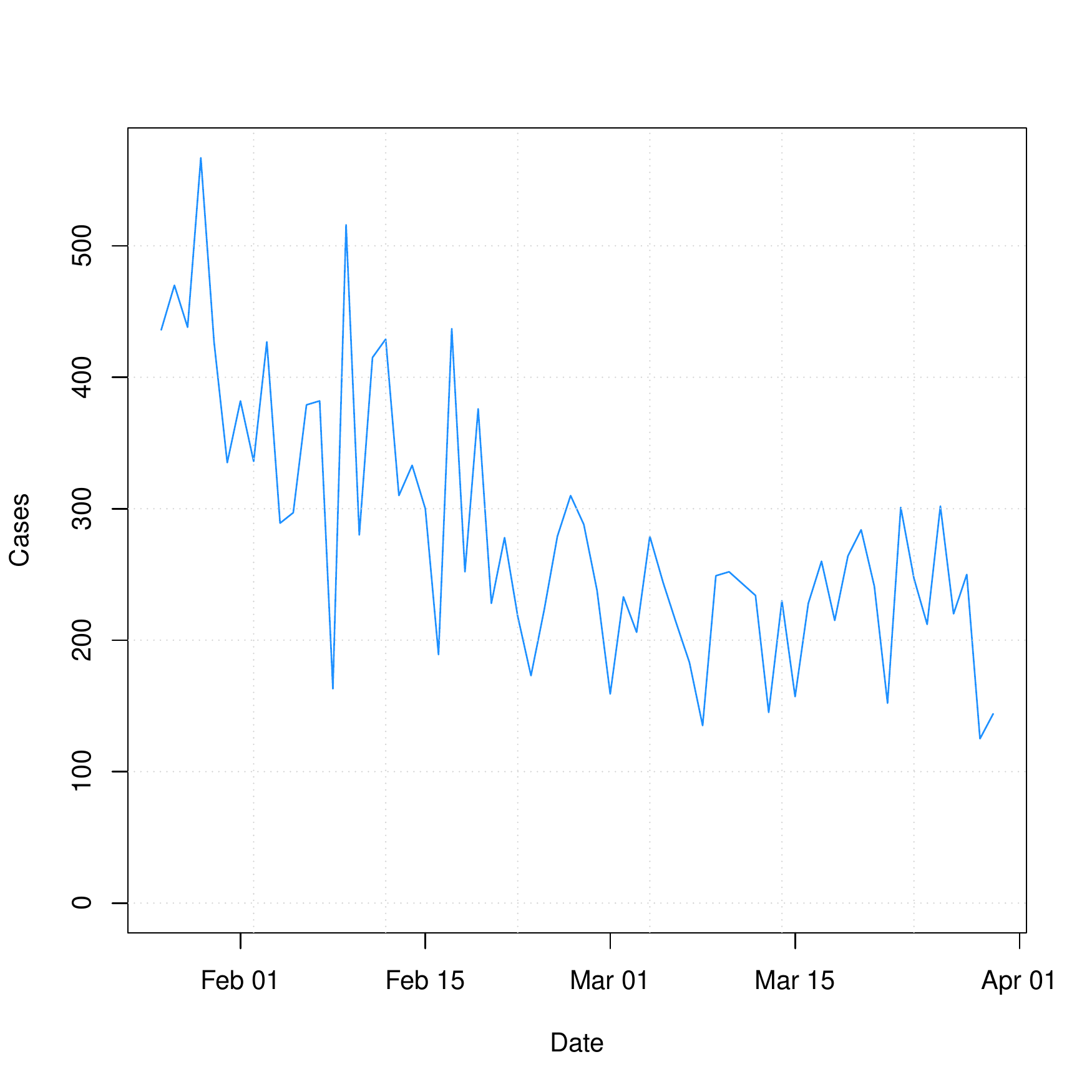}
	\end{subfigure}
	\caption{Left: Estimated economic costs for the 8 week period between 1st February 2021 and 30th March 2021. Right: Time series of actual daily cases for 8 week period between 1st February 2021 and 30th March 2021.}
	\label{fig::projected_costs}
\end{figure}

The SEIR model was projected from the 1st of February to 30th March 2021 (8 weeks) under different lockdown scenarios implemented in two 4 week chunks (the date of the second lockdown implemented was $2^\text{nd}$ March). Lockdown levels used for each four week period are specified in Table~\ref{tab::projected_deaths_and_costings}. A forward projection is made by adopting the corresponding scaling of contacts from supplementary Table~\ref{tab::interpolated_scalings}, with the SEIR system solved forward in time. Projection uncertainty bounds are obtained by taking the bootstrap replicate estimates of the corresponding scaling parameters and completing a forward projection for each of these, with $2.5\%$ and $97.5\%$ sample quantiles of these projections giving uncertainty bounds. Estimated daily economic costs are shown in the left panel of Fig~\ref{fig::projected_costs}.


\par
An estimate of the number of deaths was obtained by taking the product of the proportions from supplementary Table \ref{tab::HPSC_cases_and_deaths} and the corresponding age group from the daily removed count, estimated by
\begin{equation}
	\frac{\isuper{SI} + \isuper{SN}}{\tausub{D} - \tausub{C} + \tausub{L}} + \frac{ \isuper{PI}_i}{\tausub{D} - \tausub{C} + \tausub{L} - \tausub{R}}
\end{equation}
for each day and age group. The cost, case and death estimates are shown in Table~\ref{tab::projected_deaths_and_costings}. While the daily COVID-19 case incidence information for the Dublin region is available, the daily death data is not provided at this level of geographic granularity.

\begin{table}[ht]
	\centering
	\begin{tabular}{l|c|c|r}
		\hline
		\thead{Lockdown Scenario} & \thead{Estimated cases / 100 \\(Lower, Upper)} & \thead{Estimated deaths \\(Lower, Upper)} & \thead{Estimated cost \\ (\euro bn)} \\ 
		\hline
		No Intervention                     & 5073 (1797, 8735)    & 2785 (881, 7230)     & -12.6 \\ 
		Level 3                           & 122 (78, 307)        & 51 (34, 121)         & 1.6 \\ 
		Level 5                           & 25 (16, 72)          & 17 (12, 39)          & 11.2 \\ 
		Level 3 $\rightarrow$ Level 5     & 87 (60, 211)         & 41 (29, 92)          & 6.4 \\ 
		Level 2 $\rightarrow$ Level 5     & 118 (75, 306)        & 53 (35, 129)         & 9.1 \\ 
		\hline
	\end{tabular}
	\caption{Estimated cases, deaths and costs for the 8 week period between 1st February 2021 and $30^\text{th}$ March 2021. Where there are two scenarios listed, the shift occurred on the $2^\text{nd}$ March. The observed number of cases in the given period was 15,334 for the Dublin region. 95\% intervals are given in parentheses.}
	\label{tab::projected_deaths_and_costings}
\end{table}
\begin{figure}
	\begin{subfigure}
		\centering
		\includegraphics[scale=0.45]{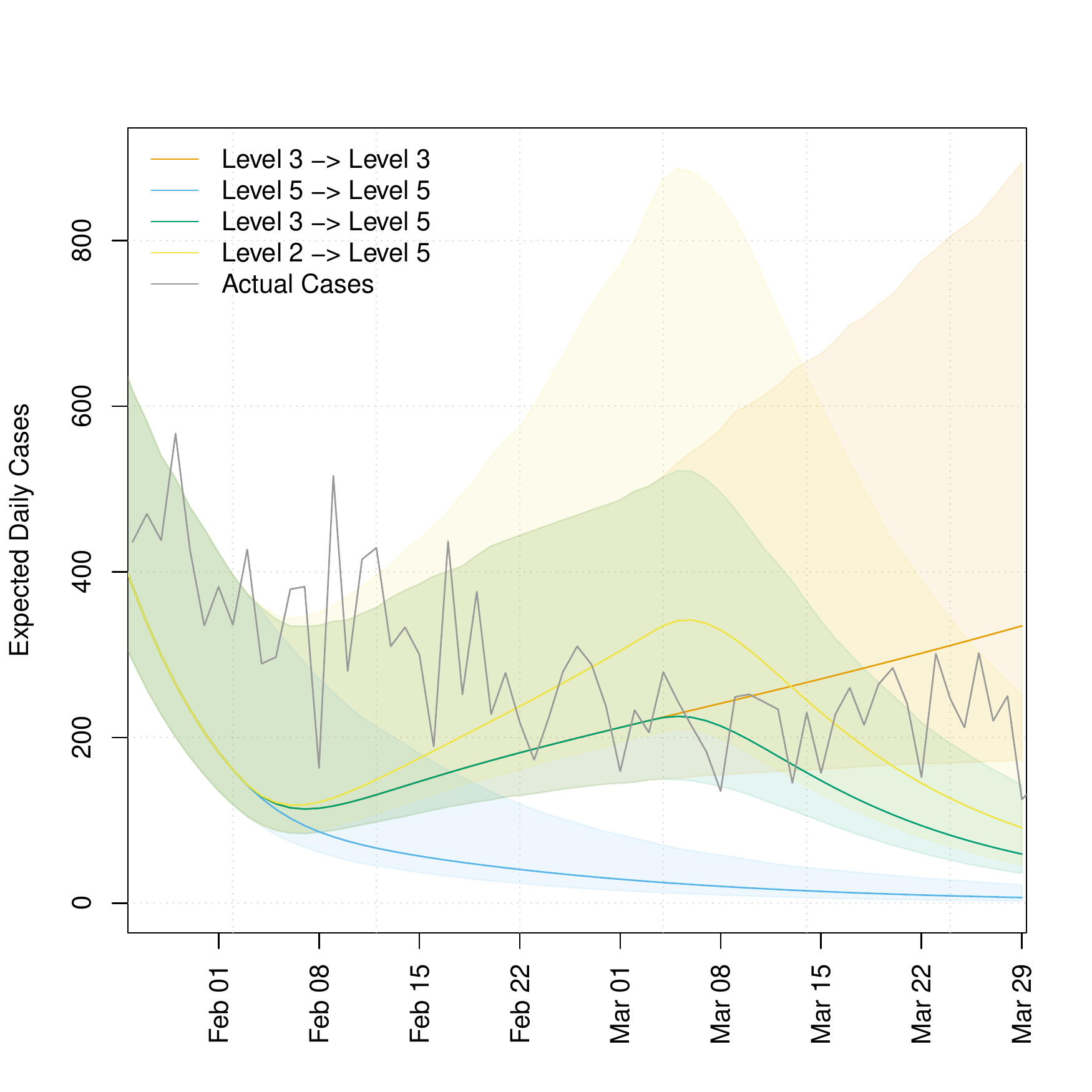}
	\end{subfigure}
	\begin{subfigure}
		\centering
		\includegraphics[scale=0.45]{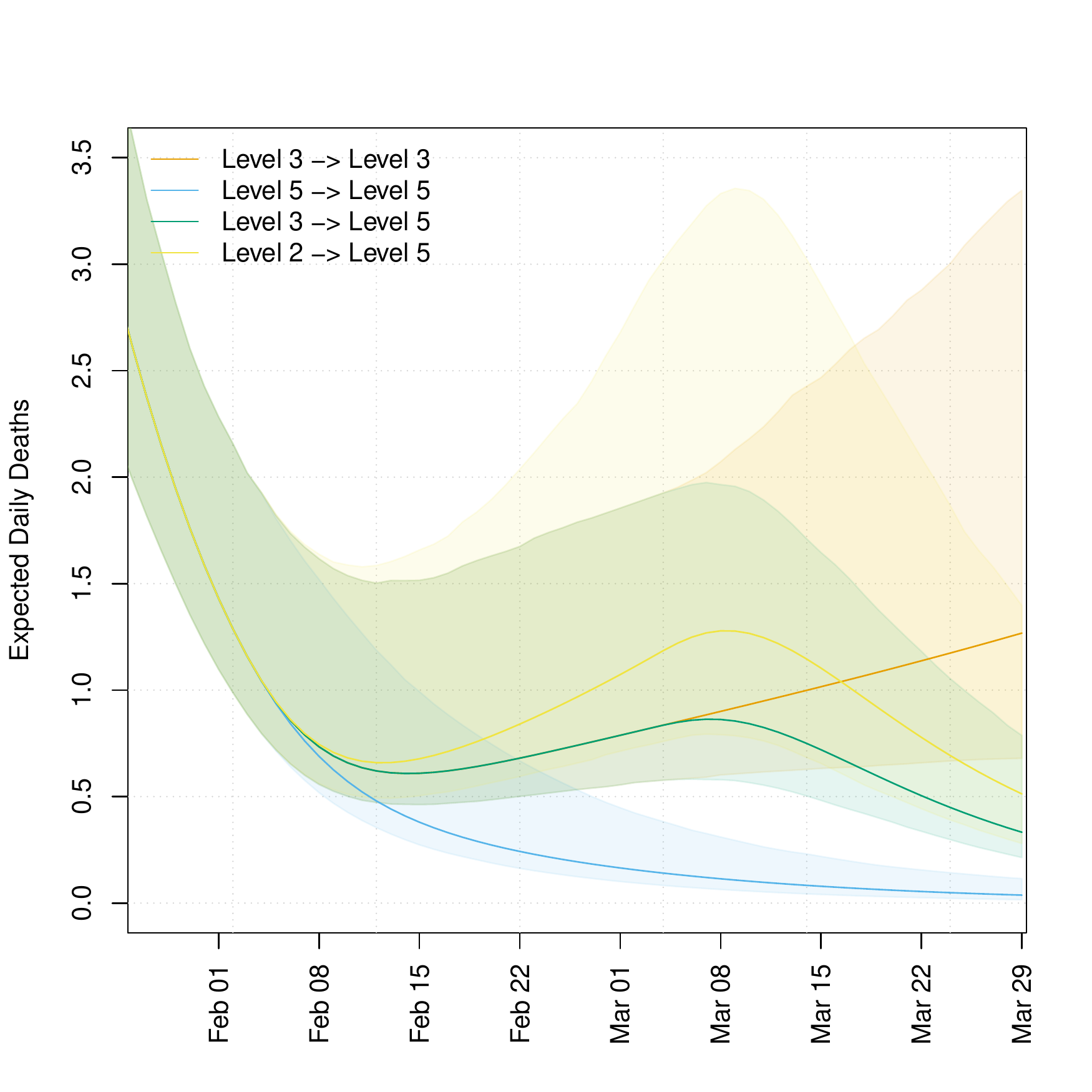}
	\end{subfigure}
	\caption{Projected Infection rates and deaths under different lockdown scenarios. The death data is not available for the Dublin region for this period.}
	\label{fig::forecast}
\end{figure}
During this time, Ireland was under Level 5 lockdown. This corresponds to our third lockdown scenario above, though as can be seen in Fig~\ref{fig::forecast}, the projection underestimates the true case count of $15,334$ that was observed during that period. In contrast, the upper bound of our estimated number of daily cases during this period is approximately $7,200$. This underestimation may be due to the public not treating this Level 5 as seriously as the Level 5 intervention last year. A closer look at the actual case counts over that period is given in the right panel of Figure~\ref{fig::projected_costs}, showing that cases plateaued after February 15th oscillating around an average figure of 200 cases per day instead of falling. A comparison across the different interventions of the total number of estimated cases for period 1st February 2021 to 30th March 2021 is given in Appendix \ref{app::tab-par-est}, Figure~\ref{fig::projected_cases}. Note: the `No Intervention' have been omitted from these graphs as the scale of the estimated cases is substantially larger than the observed cases. 


\subsection{Projecting the impact of vaccination}
\begin{figure}[ht]
	\begin{subfigure}
		\centering
		\includegraphics[scale=0.45]{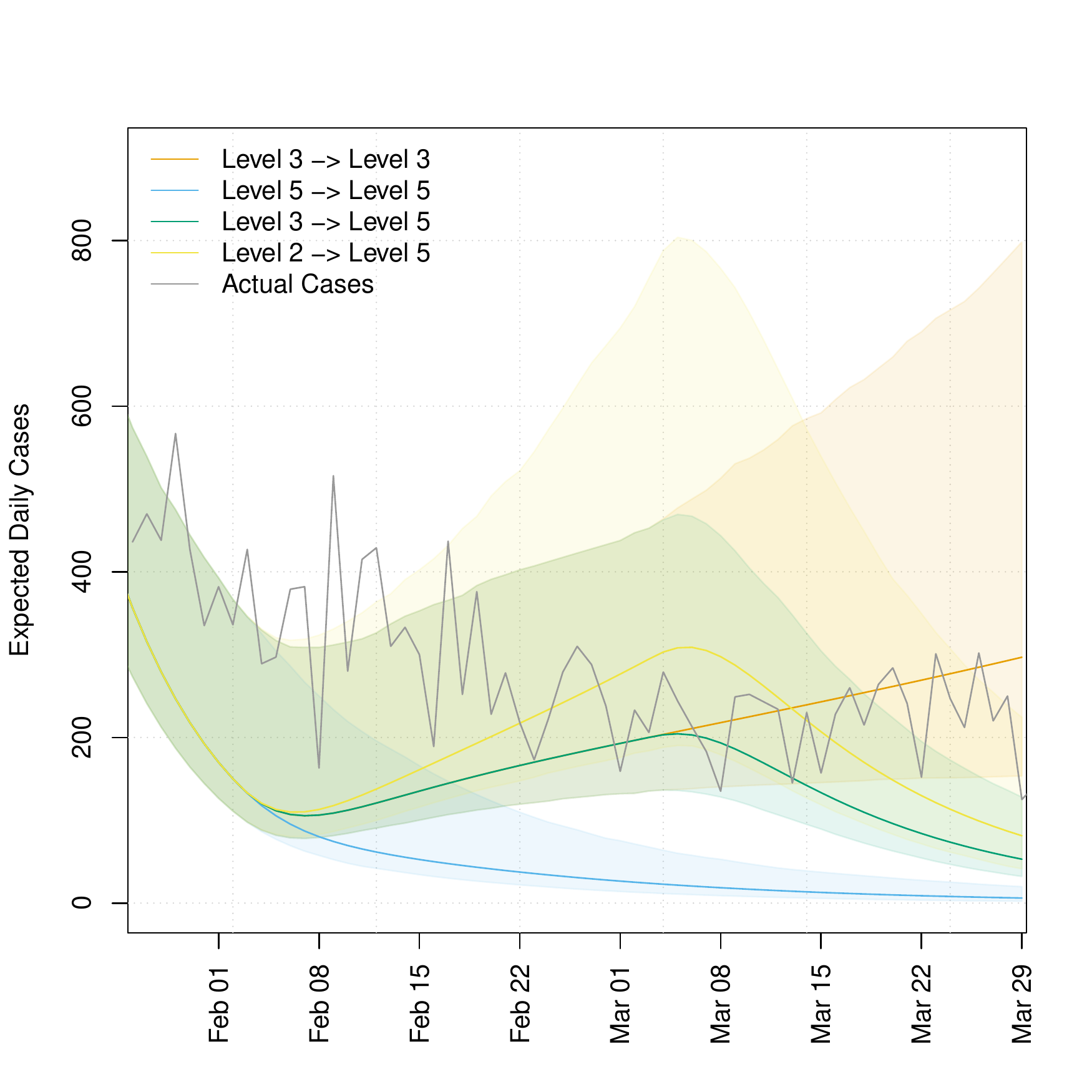}
	\end{subfigure}
	\begin{subfigure}
		\centering
		\includegraphics[scale=0.45]{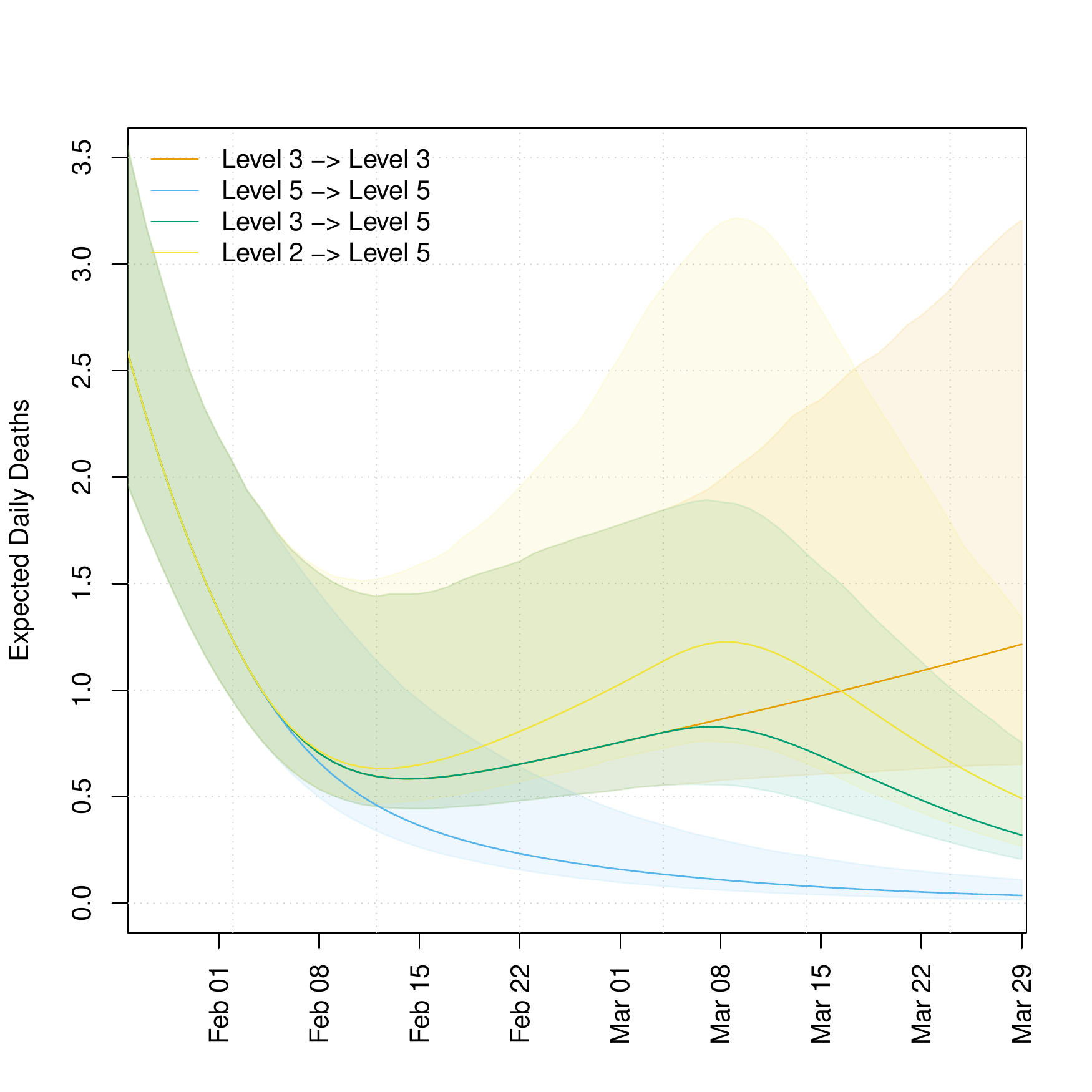}
	\end{subfigure}
	\caption{Projected Infection rates and deaths under different lockdown scenarios, where the 70+ age groups are assumed to be totally isolated post 1st December 2020 (as a proxy for vaccination).}
	\label{fig::forecast_with_isolation}
\end{figure}

As an ad-hoc proxy for the effect of vaccinations, we re-evaluated the SEIR model after setting all contacts for people aged 70 and above to zero after 1st February 2021. The symptomatic infectious rate from this model is shown in Figure~\ref{fig::forecast_with_isolation}. Estimated cases and deaths for this scenario are given in Table~\ref{tab:projections}. We estimated the 70 to 74 age class death rate by proportionally splitting the rate for 65 to 74 age class from supplementary Table~\ref{tab::HPSC_cases_and_deaths}. We see a large reduction of deaths due to the change in dynamics mimicking successful vaccination, despite the number of cases not differing substantially. This is not surprising since the 70+ age classes have considerably higher death rates.

\begin{table}[ht]
	\centering
	\begin{tabular}{l|c|c}
		\hline
		\thead{Lockdown Scenario} & \thead{Estimated Cases / 100 \\(Lower, Upper)} & \thead{Estimated Deaths \\(Lower, Upper)} \\ 
		\hline
		No Intervention                     & 5001 (1736, 8580)        & 1209 (420, 2501)     \\ 
		Level 3                           & 111 (71, 277)            & 26 (18, 62)          \\ 
		Level 5                           & 23 (15, 66)              & 9 (7, 21)            \\ 
		Level 3 $\rightarrow$ Level 5     & 79 (55, 191)             & 21 (15, 47)          \\ 
		Level 2 $\rightarrow$ Level 5     & 108 (69, 277)            & 27 (18, 66)          \\ 
		\hline
	\end{tabular}
	\caption{Estimated cases and deaths for the 8 week period between 1st February 2021 and 30th March 2021 where the contacts for everyone aged 70 and over were set to zero after 1st December 2020.}
	\label{tab:projections} 
\end{table}

\section{Discussion} \label{sec::discussion}

Our primary motivation in this article is the specification of a modelling framework that attempts to answer the question of how best public health strategies can be weighed against the attendant social and economic costs of such measures, with the recognition that these strategies can have bigger impacts on different age groups. In this regard, we have presented an age-structured SEIR model for the Irish epidemic that is generalisable to an international context. Where feasible, the parameters of our model governing disease spread have been estimated from publicly available Irish epidemiological data with a bootstrapping approach used to determine parameter uncertainties. Our fitted model captures much of the structure observed in daily case numbers. Our approach allows for local adaptions and calibrations of models in any region or location where such data is available. We also present incidence projections under a number of hypothetical government intervention strategies, in conjunction with approximate economic costings allowing for their consideration in the  decision making process. This framework is easy to interpret and suitable for describing counterfactual scenarios, which could assist policy makers with regard to minimising morbidity balanced with the costs of prospective suppression strategies. We are not aware of any framework with this level of modelling detail that has been presented to date.
\par
However, there are a number of limitations to our present approach that provide opportunities for substantial further refinements. These chiefly revolve around access to sufficiently detailed public health and economic data which would provide an ability to develop more complex models around social mixing, in addition to an ability to further refine estimates of the parameters governing disease spread and economic costings. We currently build upon the social interaction matrices provided by \citet{prem2017} which are confined to four social settings, where we model changes due to public mobility restrictions through a rescaling approach. However, with contact tracing for confirmed cases being used as a control strategy in a number of countries, access to such data would provide an avenue to substantially improve social mixing models, perhaps in conjunction with carefully constructed large scale public mobility surveys. For example, we could expand the number of social mixing venues to more than four with a better understanding on disease transmission settings. This would allow for incorporation of specific venues such as bars or restaurants, as well as allowing matrix rescalings to depend on individual age groups so that lockdown regimes have the potential to impact different age groups in different ways. 
\par
Alternatively, we could model the sociability parameters using covariates to describe the specific lockdown (e.g., schools open/closed, pubs open/closed, public events, restrictions in households etc.) rather than fixing these to have a constant value in a given lockdown level. This would allow us to make comparisons between Level 5 and Level 5+ for example -- a holiday period effect could be included as another covariate. It would also allow us to construct new hypothetical lockdown regimes.
\par
Another extension to our approach would be to incorporate uncertainty estimates around the mechanistic parameters of the SEIR model. If one could obtain reasonable uncertainty intervals on these parameters, one direction might be to use a central composite design scheme used in response surface construction \citep{Box87}, based on some transformation of these uncertainty intervals. However, such an approach would introduce a steep computational overhead and would require sufficient coding and hardware solutions to enable a feasible (in time) implementation. 

\section*{Acknowledgements}
We would like to thank Prof. James Gleeson, Prof. Brendan Murphy and Dr. Conor McAloon from the IEMAG group for their very generous help and advice. We would also like to thank Dr. David O'Sullivan and Prof. Jim Duggan for stimulating discussions. This research was funded by Science Foundation Ireland (SFI) grant number 20/COV/0166.

\footnotesize
\clearpage
\bibliographystyle{apa}
\bibliography{COVID_Research}

\begin{thebibliography}{}

\bibitem[\protect\astroncite{Beirne et~al.}{2020}]{beirne2020potential}
Beirne, K., Doorley, K., Regan, M., Roantree, B., and Tuda, D. (2020).
\newblock {The potential costs and distributional effect of Covid-19 related
  unemployment in Ireland}.
\newblock Technical report, Budget Perspectives.

\bibitem[\protect\astroncite{Box and Draper}{1987}]{Box87}
Box, G. E.~P. and Draper, N.~R. (1987).
\newblock {\em {Empirical model-building and response surfaces}}.
\newblock John Wiley and Sons.

\bibitem[\protect\astroncite{Brodeur et~al.}{2020}]{brodeur2020literature}
Brodeur, A., Gray, D.~M., Islam, A., and Bhuiyan, S. (2020).
\newblock {A Literature Review of the Economics of COVID-19}.
\newblock {\em IZA Discussion Paper}.

\bibitem[\protect\astroncite{{Buitrago-Garcia}
  et~al.}{2020}]{buitrago-garcia2020}
{Buitrago-Garcia}, D., {Egli-Gany}, D., Counotte, M.~J., Hossmann, S., Imeri,
  H., Ipekci, A.~M., Salanti, G., and Low, N. (2020).
\newblock {Occurrence and Transmission Potential of Asymptomatic and
  Presymptomatic {{SARS}}-{{CoV}}-2 Infections: {{A}} Living Systematic Review
  and Meta-Analysis}.
\newblock {\em PLOS Medicine}, 17(9):e1003346.

\bibitem[\protect\astroncite{Byrne et~al.}{2020}]{byrne2020}
Byrne, A.~W., McEvoy, D., Collins, A.~B., Hunt, K., Casey, M., Barber, A.,
  Butler, F., Griffin, J., Lane, E.~A., McAloon, C., O'Brien, K., Wall, P.,
  Walsh, K.~A., and More, S.~J. (2020).
\newblock {Inferred Duration of Infectious Period of {{SARS}}-{{CoV}}-2: Rapid
  Scoping Review and Analysis of Available Evidence for Asymptomatic and
  Symptomatic {{COVID}}-19 Cases}.
\newblock {\em BMJ Open}, 10(8):e039856.

\bibitem[\protect\astroncite{Chowell}{2017}]{chowell2017}
Chowell, G. (2017).
\newblock {Fitting Dynamic Models to Epidemic Outbreaks with Quantified
  Uncertainty: {{A}} Primer for Parameter Uncertainty, Identifiability, and
  Forecasts}.
\newblock {\em Infectious Disease Modelling}, 2(3):379--398.

\bibitem[\protect\astroncite{Coyle}{2015}]{coyle2015gdp}
Coyle, D. (2015).
\newblock {\em {GDP: a brief but affectionate history-revised and expanded
  edition}}.
\newblock Princeton University Press.

\bibitem[\protect\astroncite{{Cuevas-Maraver}
  et~al.}{2021}]{cuevas-maraver2021}
{Cuevas-Maraver}, J., Kevrekidis, P.~G., Chen, Q.~Y., Kevrekidis, G.~A.,
  {Villalobos-Daniel}, V., Rapti, Z., and Drossinos, Y. (2021).
\newblock {Lockdown Measures and Their Impact on Single- and Two-Age-Structured
  Epidemic Model for the {{COVID}}-19 Outbreak in {{Mexico}}}.
\newblock {\em Mathematical Biosciences}, page 108590.

\bibitem[\protect\astroncite{Darmody et~al.}{2020}]{Darmody20}
Darmody, M., Smith, E., and Russell, H. (2020).
\newblock {Implications of the COVID-19 pandemic for policy in relation to
  children and young people: A research review}.
\newblock {\em ESRI Survey and Statistical Report Series}.

\bibitem[\protect\astroncite{Devereux et~al.}{2020}]{devereux2020discretionary}
Devereux, M.~P., G{\"u}{\c{c}}eri, {\.I}., Simmler, M., and Tam, E.~H. (2020).
\newblock {Discretionary fiscal responses to the COVID-19 pandemic}.
\newblock {\em Oxford Review of Economic Policy}, 36(Supplement\_1):S225--S241.

\bibitem[\protect\astroncite{Diekmann et~al.}{1990}]{diekmann1990}
Diekmann, O., Heesterbeek, J. A.~P., and Metz, J. A.~J. (1990).
\newblock {On the Definition and the Computation of the Basic Reproduction
  Ratio {{R0}} in Models for Infectious Diseases in Heterogeneous Populations}.
\newblock {\em Journal of Mathematical Biology}, 28(4):365--382.

\bibitem[\protect\astroncite{Eichenbaum
  et~al.}{2020}]{eichenbaum2020macroeconomics}
Eichenbaum, M.~S., Rebelo, S., and Trabandt, M. (2020).
\newblock The macroeconomics of epidemics.
\newblock Technical report, National Bureau of Economic Research.

\bibitem[\protect\astroncite{Evoy et~al.}{2020}]{evoy2020}
Evoy, D.~M., McAloon, C.~G., Collins, A.~B., Hunt, K., Butler, F., Byrne,
  A.~W., Casey, M., Barber, A., Griffin, J.~M., Lane, E.~A., Wall, P., and
  More, S.~J. (2020).
\newblock {The Relative Infectiousness of Asymptomatic {{SARS}}-{{CoV}}-2
  Infected Persons Compared with Symptomatic Individuals: {{A}} Rapid Scoping
  Review}.
\newblock {\em medRxiv}, page 2020.07.30.20165084.

\bibitem[\protect\astroncite{Fitzgerald}{2020}]{fitzgerald2020national}
Fitzgerald, J. (2020).
\newblock {National accounts for a global economy: the case of Ireland}.
\newblock In {\em The Challenges of Globalization in the Measurement of
  National Accounts}. University of Chicago Press.

\bibitem[\protect\astroncite{Fumanelli et~al.}{2012}]{fumanelli2012}
Fumanelli, L., Ajelli, M., Manfredi, P., Vespignani, A., and Merler, S. (2012).
\newblock {Inferring the structure of social contacts from demographic data in
  the analysis of infectious diseases spread}.
\newblock {\em PLoS Comput Biol}, 8(9):e1002673.

\bibitem[\protect\astroncite{Griffin et~al.}{2020}]{griffin2020rapid}
Griffin, J., Casey, M., Collins, {\'A}., Hunt, K., McEvoy, D., Byrne, A.,
  McAloon, C., Barber, A., Lane, E.~A., and More, S. (2020).
\newblock {Rapid review of available evidence on the serial interval and
  generation time of COVID-19}.
\newblock {\em BMJ open}, 10(11):e040263.

\bibitem[\protect\astroncite{Grimm et~al.}{2021}]{grimm2021}
Grimm, V., Mengel, F., and Schmidt, M. (2021).
\newblock {Extensions of the {{SEIR}} Model for the Analysis of Tailored Social
  Distancing and Tracing Approaches to Cope with {{COVID}}-19}.
\newblock {\em Scientific Reports}, 11(1):4214.

\bibitem[\protect\astroncite{Heffernan et~al.}{2005}]{heffernan2005}
Heffernan, J., Smith, R., and Wahl, L. (2005).
\newblock {Perspectives on the Basic Reproductive Ratio}.
\newblock {\em Journal of the Royal Society Interface}, 2(4):281--293.

\bibitem[\protect\astroncite{HPSC}{2020}]{HPSC2020}
HPSC (2020).
\newblock {Epidemiology of {{COVID}}-19 in {{Ireland}}- Daily Reports,
  {{December}} 2020}.
\newblock
  https://www.hpsc.ie/a-z/respiratory/coronavirus/novelcoronavirus/casesinireland/epidemiologyofcovid-19inireland/december2020/COVID-19
  Daily epidemiology report (NPHET)\_20201230\_website.pdf.

\bibitem[\protect\astroncite{IEMAG}{2020}]{iemag2020}
IEMAG (2020).
\newblock {A Population-Level {{SEIR}} Model for {{COVID}}-19 Scenarios}.
\newblock Technical report, {Irish Department of Health}.

\bibitem[\protect\astroncite{Kaplan et~al.}{2020}]{kaplan2020great}
Kaplan, G., Moll, B., and Violante, G.~L. (2020).
\newblock {The great lockdown and the big stimulus: Tracing the pandemic
  possibility frontier for the US}.
\newblock Technical report, National Bureau of Economic Research.

\bibitem[\protect\astroncite{Kimathi et~al.}{2021}]{kimathi2021}
Kimathi, M., Mwalili, S., Ojiambo, V., and Gathungu, D.~K. (2021).
\newblock {Age-Structured Model for {{COVID}}-19: {{Effectiveness}} of Social
  Distancing and Contact Reduction in {{Kenya}}}.
\newblock {\em Infectious Disease Modelling}, 6:15--23.

\bibitem[\protect\astroncite{Kwong et~al.}{2020}]{Kwong20}
Kwong, A. S.~F., Pearson, R.~M., Adams, M.~J., Northstone, K., Tilling, K.,
  Smith, D., Fawns-Ritchie, C., Bould, H., Warne, N., Zammit, S., and et~al.
  (2020).
\newblock {Mental health before and during the COVID-19 pandemic in two
  longitudinal UK population cohorts}.
\newblock {\em The British Journal of Psychiatry}, page 1–10.

\bibitem[\protect\astroncite{Lee et~al.}{2021}]{lee2021}
Lee, T., Kwon, H.-D., and Lee, J. (2021).
\newblock {The Effect of Control Measures on {{COVID}}-19 Transmission in
  {{South Korea}}}.
\newblock {\em PLOS ONE}, 16(3):e0249262.

\bibitem[\protect\astroncite{McAloon et~al.}{2020}]{mcaloon2020}
McAloon, C., Collins, {\'A}., Hunt, K., Barber, A., Byrne, A.~W., Butler, F.,
  Casey, M., Griffin, J., Lane, E., McEvoy, D., Wall, P., Green, M., O'Grady,
  L., and More, S.~J. (2020).
\newblock {Incubation Period of {{COVID}}-19: A Rapid Systematic Review and
  Meta-Analysis of Observational Research}.
\newblock {\em BMJ Open}, 10(8):e039652.

\bibitem[\protect\astroncite{Mossong et~al.}{2008}]{mossong2008}
Mossong, J., Hens, N., Jit, M., Beutels, P., Auranen, K., Mikolajczyk, R.,
  Massari, M., Salmaso, S., Tomba, G.~S., Wallinga, J., Heijne, J.,
  {Sadkowska-Todys}, M., Rosinska, M., and Edmunds, W.~J. (2008).
\newblock {Social {{Contacts}} and {{Mixing Patterns Relevant}} to the
  {{Spread}} of {{Infectious Diseases}}}.
\newblock {\em PLOS Medicine}, 5(3):e74.

\bibitem[\protect\astroncite{N{\'a}raigh and Byrne}{2020}]{naraigh2020}
N{\'a}raigh, L.~{\'O}. and Byrne, {\'A}. (2020).
\newblock {Piecewise-Constant Optimal Control Strategies for Controlling the
  Outbreak of {{COVID}}-19 in the {{Irish}} Population}.
\newblock {\em Mathematical Biosciences}, 330:108496.

\bibitem[\protect\astroncite{Nash et~al.}{2011}]{nash2011unifying}
Nash, J.~C., Varadhan, R., et~al. (2011).
\newblock {Unifying optimization algorithms to aid software system users:
  optimx for R}.
\newblock {\em Journal of Statistical Software}, 43(9):1--14.

\bibitem[\protect\astroncite{Nelder and Mead}{1965}]{nelder1965simplex}
Nelder, J.~A. and Mead, R. (1965).
\newblock A simplex method for function minimization.
\newblock {\em The computer journal}, 7(4):308--313.

\bibitem[\protect\astroncite{Nishiura and Chowell}{2009}]{nishiura2009}
Nishiura, H. and Chowell, G. (2009).
\newblock {The {{Effective Reproduction Number}} as a {{Prelude}} to
  {{Statistical Estimation}} of {{Time}}-{{Dependent Epidemic Trends}}}.
\newblock {\em Mathematical and Statistical Estimation Approaches in
  Epidemiology}, pages 103--121.

\bibitem[\protect\astroncite{Pawitan}{2001}]{Pawitan05}
Pawitan, Y. (2001).
\newblock {\em {In all likelihood}}.
\newblock Oxford University Press.

\bibitem[\protect\astroncite{Prem et~al.}{2017}]{prem2017}
Prem, K., Cook, A.~R., and Jit, M. (2017).
\newblock {Projecting Social Contact Matrices in 152 Countries Using Contact
  Surveys and Demographic Data}.
\newblock {\em PLOS Computational Biology}, 13(9):e1005697.

\bibitem[\protect\astroncite{Prem et~al.}{2020}]{prem2020}
Prem, K., Liu, Y., Russell, T.~W., Kucharski, A.~J., Eggo, R.~M., Davies, N.,
  Jit, M., Klepac, P., Flasche, S., Clifford, S., Pearson, C. A.~B., Munday,
  J.~D., Abbott, S., Gibbs, H., Rosello, A., Quilty, B.~J., Jombart, T., Sun,
  F., Diamond, C., Gimma, A., {van Zandvoort}, K., Funk, S., Jarvis, C.~I.,
  Edmunds, W.~J., Bosse, N.~I., and Hellewell, J. (2020).
\newblock {The Effect of Control Strategies to Reduce Social Mixing on Outcomes
  of the {{COVID}}-19 Epidemic in {{Wuhan}}, {{China}}: A Modelling Study}.
\newblock {\em The Lancet Public Health}, 5(5):e261--e270.

\bibitem[\protect\astroncite{Soetaert et~al.}{2010}]{soetaert2010solving}
Soetaert, K., Petzoldt, T., and Setzer, R.~W. (2010).
\newblock {Solving Differential Equations in {R}: Package de{S}olve}.
\newblock {\em Journal of Statistical Software}, 33(9):1--25.

\bibitem[\protect\astroncite{Taylor}{2021}]{Taylorn879}
Taylor, L. (2021).
\newblock {Covid-19: Brazil{\textquoteright}s spiralling crisis is increasingly
  affecting young people}.
\newblock {\em BMJ}, 373.

\bibitem[\protect\astroncite{Teimouri}{2020}]{teimouri2020}
Teimouri, A. (2020).
\newblock {An {{SEIR Model}} with {{Contact Tracing}} and {{Age}}-{{Structured
  Social Mixing}} for {{COVID}}-19 Outbreak}.
\newblock {\em medRxiv}, page 2020.07.05.20146647.

\bibitem[\protect\astroncite{Woloszko}{2020}]{wolosko:2020}
Woloszko, N. (2020).
\newblock Tracking activity in real time with google trends.
\newblock {\em OECD Economics Department Working Papers}, 1634.

\end{thebibliography}

\pagebreak

\setcounter{table}{0}
\renewcommand{\thetable}{S\arabic{table}}%
\setcounter{figure}{0}
\renewcommand{\thefigure}{S\arabic{figure}}%

\section*{Supplementary material}


\subsection*{A: Lockdown measures in Ireland} \label{app::tab_lockdown}

A more detailed overview of the government interventions over the course of the epidemic in Ireland from what we define as day 0 of the epidemic, 28th February 2020, up to and including 31st January 2021. We acknowledge that certain interventions were not eased for the whole indicated period but for the purpose of simplification we are assuming they were.

\begin{table}[h]
	\centering
	\begin{tabular}{||ll p{10cm}||}
		\hline
		Start      & End        & Government Intervention  \\
		\hline
		\hline
		29/02/2020 & 11/03/2020 & No Intervention      \\
		\hline
		12/03/2020 & 26/03/2020 & All schools and universities closed, followed by bars and a ban on mass gathering (although there was a four day delay between the schools closure and the closure of the bars, we decided to treat it as the same period for simplification)\\
		\hline
		27/03/2020 & 18/05/2020 & A strict lockdown whereby all non-essential services and industries were put on hold \\
		\hline
		19/05/2020 & 07/06/2020 & The first phase of the gradual easing of the strict lockdown \\
		\hline
		08/06/2020 & 28/06/2020 & The second phase of the gradual easing of the strict lockdown\\
		\hline
		29/06/2020 & 17/08/2020 & The third phase of the gradual easing of the strict lockdown \\
		\hline
		18/08/2020 & 17/09/2020 & Some level of restrictions was brought back, although the government's plan of living with COVID--19 and the introduction of the 5 levels lockdown tiers was introduced later during this phase, the restrictions were very similar to those of what the government now call a Level 2 lockdown, hence we label the intervention implemented during this phase as a Level 2 lockdown.\\
		\hline
		18/09/2020 & 20/10/2020 & Level 3 lockdown\\
		\hline
		21/10/2020 & 30/11/2020 & Level 5 lockdown   \\
		\hline
		01/12/2020 & 18/12/2020 & Level 3 lockdown, all non-essential retail, hairdressers, gyms, leisure centres, museums, galleries, libraries, cinemas and places of worship were allowed to reopen. Restaurants, gastropubs and hotel restaurants were allowed to provide indoor service.\\
		\hline
		19/12/2020 & 03/01/2021 & Holiday period, households were allowed to mix with up to two other households and inter-county travel was permitted. \\
		\hline
		04/01/2021 & 30/01/2021 & Level 5 lockdown, plus the closure of schools and the construction industry.\\
		\hline
	\end{tabular}
	\caption{A summary of the government interventions that were implemented over the study time frame. \label{tab::policy_summary}}
\end{table}

\clearpage
\subsection*{B  Next generation matrix} \label{app::next-gen}

\newcommand{\bF}{\mathbf{F}}


The next generation matrix encodes the spread of the disease as a linear operator whose form is determined by the model.\citet{diekmann1990} showed that (subject to light conditions) the dominant eigenvalue of the \emph{next-generation operator} can be interpreted as ``the typical number of secondary cases'', or $R_0$. For \emph{discrete} state models such as ours, Section 2.2 of \citet{heffernan2005} provides a practical explanation of how to construct the next generation \emph{matrix}. We go through the method here for our model.
\par
Let $\tilde{\mathbf{z}} = (\tilde{z}_1, \dots, \tilde{z}_p)$ be the vector of compartment sizes for compartments from which infected individuals enter or leave (i.e, all except the susceptible and removed compartments). The entries of $\tilde{\mathbf{z}}$ are extracted from $\mathbf{z}$ as defined in Section 3 of the manuscript. Now introduce $f_i(\tilde{\mathbf{z}})$ as the rate of \emph{new} infections that enter compartment $i$, let $v_i^+(\tilde{\mathbf{z}})$ be the rate of individuals arriving into compartment $i$ who are \emph{not} newly infected, let $v_i^-(\tilde{\mathbf{z}})$ be the rate of individuals leaving compartment $i$, and finally let $v_i(\tilde{\mathbf{z}}) = v_i^-(\tilde{\mathbf{z}}) - v_i^+(\tilde{\mathbf{z}})$. With this notation, every system of equations described in section \ref{sec::epi_model} can be expressed as $f_i(\tilde{\mathbf{z}}) - v_i(\tilde{\mathbf{z}})$.\\
The next generation matrix is constructed from the matrices of partial derivatives of $f_i$ and $v_i$,
\begin{equation}
	F_{ij} = \frac{\partial f_i}{\partial \tilde{z}_j}(\tilde{\mathbf{z}}_0), \quad \quad V_{ij} = \frac{\partial v_i}{\partial \tilde{z}_j}(\tilde{\mathbf{z}}_0)
\end{equation}
evaluated at the \emph{disease-free equilibrium}, $\tilde{\mathbf{z}}_0$, i.e., the point at which no infection is present. In our application, since we are imposing a constant population $N$, the disease free equilibrium simply means that $S_i = N_i, i=1,\dots,A$ and all other compartments equal 0. The next generation matrix $\mathbf{Q}$ is equal to
\begin{equation}
	\mathbf{Q} = \bF\, \mathbf{V}^{-1}
\end{equation}
In our application, the $\bF$ matrix can be expressed as a block matrix, where each block corresponds to an age group

\begin{center}
	$$ \bF =
	\begin{bmatrix}
		\bF_{11} & \bF_{12} & . & . & . & \bF_{1A}\\
		\bF_{21} & \bF_{22} & . & . & . & \bF_{2A}\\
		.&.&. & & &.\\
		.&.& &.&&.\\
		.&.& & &.&.\\
		\bF_{A1} & \bF_{A2} & . & . & . & \bF_{AA}
	\end{bmatrix}
	$$
\end{center}

where each block is given by

\begin{center}
	$$
	\bF_{mn} = 
	\begin{bmatrix}
		0 & \alpha B_{mn} & B_{mn} & \kappa B_{mn} & B_{mn} & \kappa B_{mn} & B_{mn}\\
		0 & 0 & 0 & 0 & 0 & 0 & 0\\
		0 & 0 & 0 & 0 & 0 & 0 & 0\\
		0 & 0 & 0 & 0 & 0 & 0 & 0\\
		0 & 0 & 0 & 0 & 0 & 0 & 0\\
		0 & 0 & 0 & 0 & 0 & 0 & 0\\
		0 & 0 & 0 & 0 & 0 & 0 & 0\\
	\end{bmatrix}.
	$$
\end{center}
Note that only the first row of each block is non-zero and $B_{mn}$ is defined as follows
\begin{equation*}
	B_{mn} = \beta c_{mn}\frac{N_m}{N_n}.
\end{equation*}
Beware that the subscripts $m$ and $n$ here correspond to the \emph{block} indices, not its cell position in the matrix. Assuming there is no movement between the age groups the $\mathbf{V}$ matrix is expressed as a block diagonal matrix where each block corresponds to an age group.

\begin{center}
	$$ \bV =
	\begin{bmatrix}
		\bV_{11} & 0 & . & . & . & 0\\
		0 & \bV_{22}& . & . & . & 0\\
		.&.&. & & &.\\
		.&.& &.&&.\\
		.&.& & &.&.\\
		0& 0 & . & . & . & \bV_{AA}
	\end{bmatrix}
	$$
\end{center}Each block is given by
\begin{center}
	$$
	\bV_{ii} = 
	\begin{bmatrix}
		\frac{1}{\tausub{L}} & 0 & 0 & 0 & 0 & 0 & 0\\
		-\frac{\fsub{AS}}{\tausub{L}} & \frac{1}{\tausub{D}} & 0 & 0 & 0 & 0 & 0\\
		-\frac{1-\fsub{AS}}{\tausub{L}} & 0 & \frac{1}{\tausub{C} - \tausub{L}} & 0 & 0 & 0 & 0\\
		0 & 0 & -\frac{\fsub{SI}}{\tausub{C} - \tausub{L}} & \frac{1}{\tausub{D} - \tausub{C} - \tausub{L}} & 0 & 0 & 0\\
		0 & 0 & -\frac{\fsub{T}}{\tausub{C} - \tausub{L}} & 0 & \frac{1}{\tausub{R}} & 0 & 0 \\
		0 & 0 & -\frac{1-\fsub{SI}-\fsub{T}}{\tausub{C}-\tausub{L}} & 0 & 0 & \frac{1}{\tausub{C}-\tausub{D}+\tausub{L}} & 0\\
		0 & 0 & 0 & 0 & -\frac{1}{\tausub{R}} & 0 & \frac{1}{\tausub{D}-\tausub{C}+\tausub{L}-\tausub{R}}
	\end{bmatrix}
	$$
\end{center}
Since each block represents an age group, different parameters for different age can be easily incorporated by simply altering the suitable block.\\

The largest absolute eigenvalue value of $\mathbf{Q}$ is $R_0$. Since $\beta$ is easily factored out of the $\bF$ matrix, the eigenvalue can be expressed as a product of $\beta$ and the maximum eigenvalue of $\widehat{\bF}\, \bV^{-1}$, where $\bF = \beta\, \widehat{\bF}$. This means that if $R_0$ is determined and $\beta$ is desired, the expression can easily be re-arranged:
\begin{equation}
	\beta = R_0/\xi
\end{equation}
where $\xi$ is the maximum eigenvalue of $\widehat{\bF}\,\bV^{-1}$. 

\newpage

\subsection*{C Additional Tables and Figures } \label{app::tab-par-est}

\begin{table}[ht]
	\centering
	\begin{tabular}{||c  p{5cm} c p{5cm}||} 
		\hline
		Parameter & Description & Value & Reference\\ [0.5ex] 
		\hline\hline
		$\tausub{C}$ & Average incubation period & 5.8 & \cite{mcaloon2020}\\ 
		\hline
		$\tausub{P}$ & Average pre-symptomatic period & 2 &  \cite{byrne2020}\\
		\hline
		$\tausub{L}$ & Average latent period. This is computed as $\tausub{C}$ minus $\tausub{P}$ & 3.8 & \\
		\hline
		$\tausub{D}^C$ & Average infectious period for symptomatic patients & 13.4 &  \cite{byrne2020}\\
		\hline
		$\tausub{D}^{SC}$ & Average infectious period for asymptomatic patients & 6 &  \cite{byrne2020}\\
		\hline
		$\tausub{D}$ & Average infectious period. Weighted average of the symptomatic and asymptomatic periods (weighted by prevalence) & 13.5 & \\
		\hline
		$R_0$ & Basic reproductive number & 3.4 & \cite{naraigh2020} \\
		\hline
		$\alpha$ & Factor reduction of transmission from asymptomatic cases& 0.55 & \cite{evoy2020} (The mean of given interval) \\
		\hline
		$\kappa$ & Factor reduction of transmission from isolating cases & 0.05 & \cite{iemag2020} \\
		\hline
		$\fsub{AS}$ & Proportion of asymptomatic infections & 0.20  & \cite{buitrago-garcia2020}\\
		\hline
		$\fsub{T}$ & Proportion symptomatic who get tested & 0.8 & \cite{iemag2020}  \\
		\hline
		$\fsub{SI}$ & Proportion symptomatic who self-isolate & 0.1 & \cite{iemag2020} \\
		\hline
		$\tausub{R}$ & Expected time between first symptoms and test result & 7 & \cite{iemag2020}\\
		\hline 
	\end{tabular} 
	\caption{List of parameters used directly in or sourced to for the specification of the SEIR model. }
	\label{tab::par}
\end{table}


\begin{table}[ht]
	\centering
	\begin{tabular}{||lrrrr||}
		\hline
		Policy & $\widehat{\boldtheta}$ & Median & Lower Bound & Upper Bound \\ 
		\hline
		\hline
		No Intervention & 1.270 & 1.342 &         0.904 & 1.935 \\ 
		School Closure & 2.312 & 2.244 &          1.777 & 2.722 \\ 
		Intense Lockdown & 0.176 & 0.175 &        0.148 & 0.205 \\ 
		Relax Intervention 1 & 0.003 & 0.004 &    $<0.001$ & 0.021 \\ 
		Relax Intervention 2 & 0.169 & 0.187 &    0.100 & 0.293 \\ 
		Relax Intervention 3 & 0.485 & 0.470 &    0.399 & 0.546 \\ 
		Lockdown Level 2 & 0.466 & 0.471 &        0.388 & 0.557 \\ 
		Lockdown Level 3 & 0.398 & 0.398 &        0.339 & 0.462 \\ 
		Lockdown Level 5 & 0.118 & 0.117 &        0.072 & 0.165 \\ 
		Lockdown Level 3+ & 1.041 & 1.046 &       0.761 & 1.408 \\ 
		Holiday period & 0.829 & 0.812 &          0.589 & 1.099 \\ 
		Lockdown Level 5+  & 0.015 & 0.016 &      $<0.001$ & 0.124 \\ 
		\hline
	\end{tabular}
	\caption{The estimated contact matrix scaling parameters for each \textbf{observed} lockdown phase, with bootstrapped $95\%$ uncertainty bounds. \label{tab::est_lockdown_scalings}}
\end{table}





\begin{table}[ht]
	\centering
	\begin{tabular}{||lrrrr||}
		\hline
		Policy & Scalars & Median & Lower Bound & Upper Bound \\ 
		\hline
		\hline
		Lockdown Level 0 & 1.270 & 1.342 & 0.904 & 1.935 \\ 
		Lockdown Level 1 & 0.868 & 0.908 & 0.691 & 1.200 \\ 
		Lockdown Level 2 & 0.466 & 0.471 & 0.388 & 0.557 \\ 
		Lockdown Level 3 & 0.398 & 0.398 & 0.339 & 0.462 \\ 
		Lockdown Level 4 & 0.258 & 0.258 & 0.232 & 0.286 \\ 
		Lockdown Level 5 & 0.118 & 0.117 & 0.072 & 0.165 \\ 
		\hline
	\end{tabular}
	\caption{The estimated scaling parameters for each of the lockdown levels specified by the Irish Government.}
	\label{tab::interpolated_scalings}
\end{table}



\begin{table}[ht]
	\centering
	\begin{tabular}{||c | c | c | c ||} 
		\hline
		Age Range & Total Cases & Total Deaths & Proportion of Deaths to Cases\\
		\hline
		\hline
		0-14 & 8552 & 0 & 0\\
		\hline
		15-24 & 15505 & $<5$ & $<3.2\times10^{-4}$\\
		\hline
		25-34 & 15120 & 6 & $3.9\times10^{-4}$\\
		\hline
		35-44 & 13900 & 14 & $10^{-3}$\\
		\hline
		45-54 & 13176 & 38 & $2.9\times10^{-3}$\\
		\hline
		55-64 & 9445 & 94 & 0.01 \\
		\hline
		65-74 & 4977 & 332 & 0.07 \\ 
		\hline
		75+ & 7734 & 1736 & 0.22 \\
		\hline
	\end{tabular} 
	\caption{HPSC's COVID--19 total case and death rate in Ireland up to 28th December 2020 (this was the latest point that the HPSC reported the full cumulative death rates with these age groups). Where there were more than 0 but less than 5 cases in a cell, the HPSC did not report the number to retain anonymity. There were a further 30 cases where the age was not known.}
	\label{tab::HPSC_cases_and_deaths}
\end{table}

\newpage

\begin{figure}[ht]
	\centering
	\includegraphics[scale=0.7]{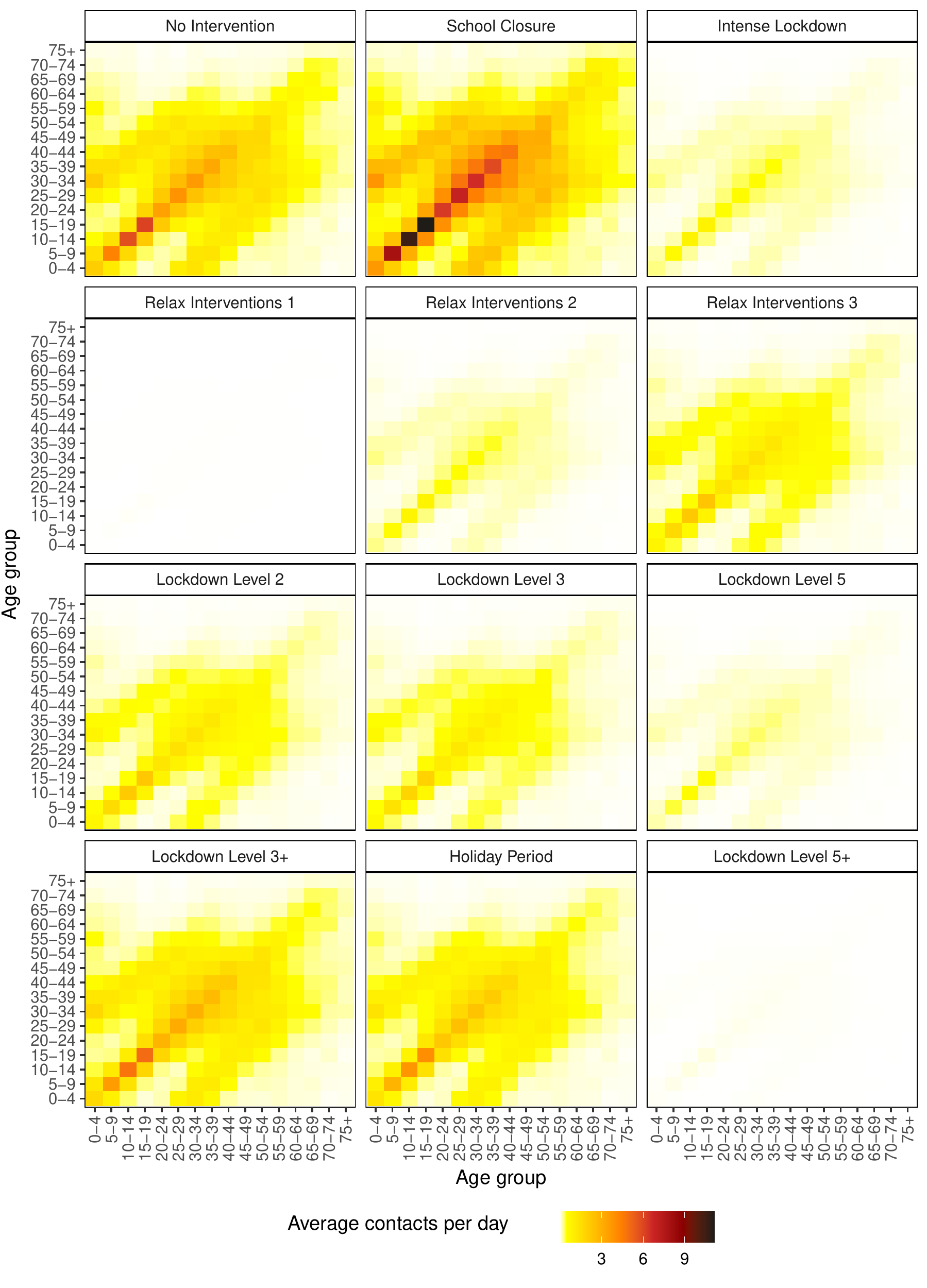}
	\caption{The estimated contact matrix for each of the policies witnessed over the study time frame.}
	\label{fig::scaled_cm}
\end{figure}

\begin{figure} [ht]
	\centering
	\includegraphics[scale=0.7]{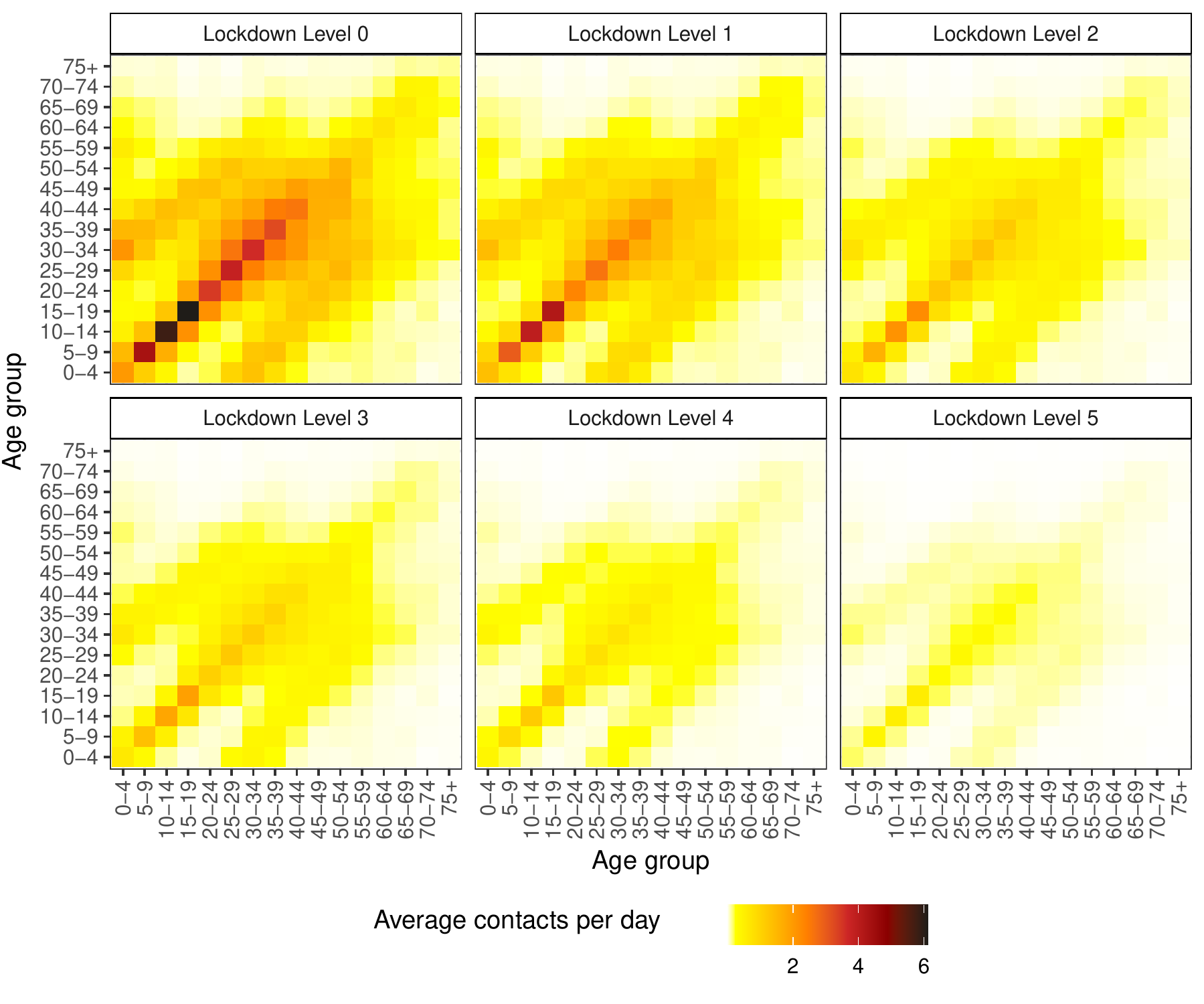}
	\caption{Estimated contacts matrix for each of the 5 lockdown levels specified by the Irish Government.}
	\label{fig::contacts_heatmap_inter}
\end{figure}



\begin{figure}[ht]
	\centering
	\includegraphics[scale=0.7]{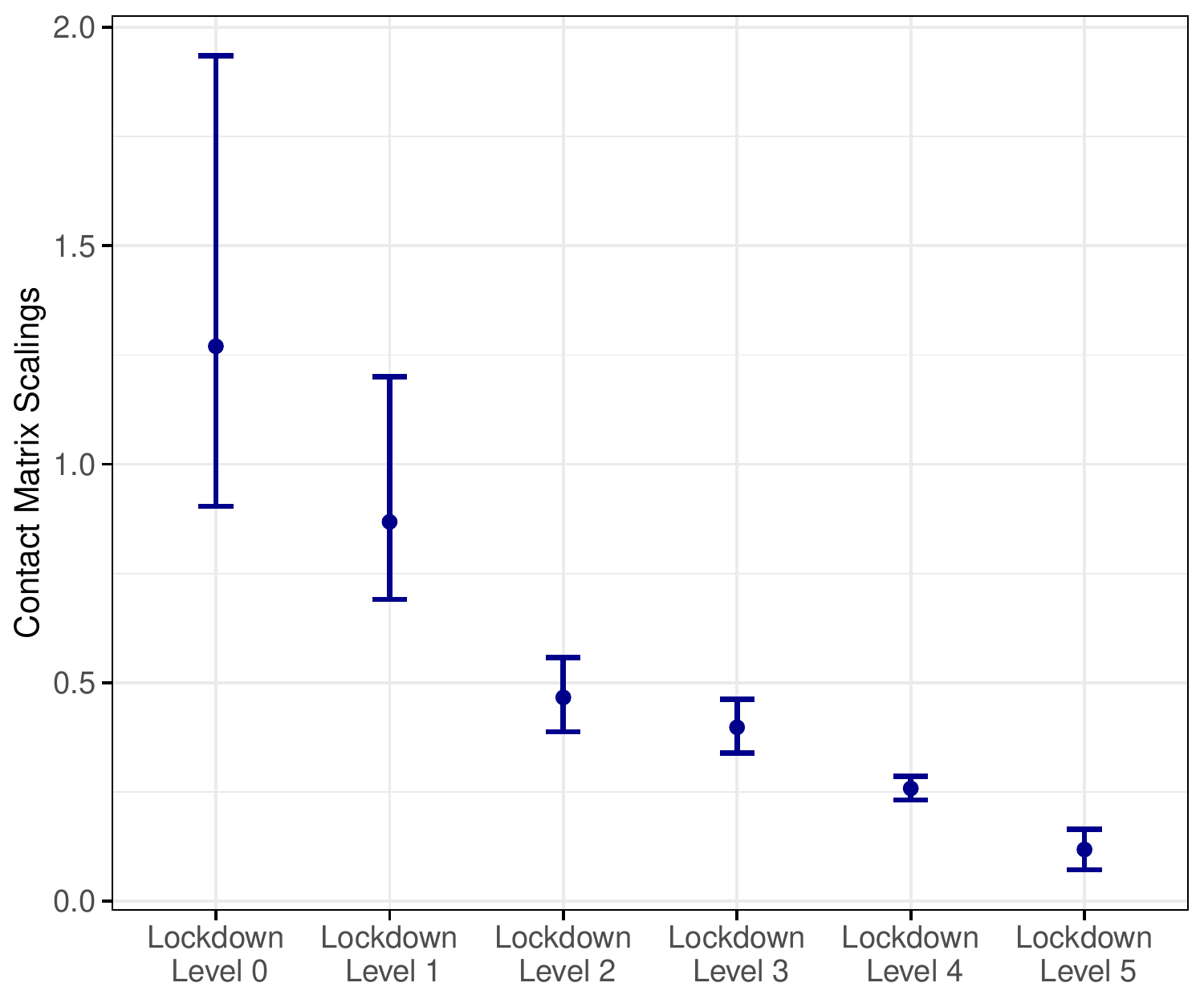}
	\caption{The estimated contact matrix scaling parameters for Government specified projections.}
	\label{fig::cm_scales_2}
\end{figure}

\begin{figure}[ht]
	\centering
	\includegraphics[scale=0.57]{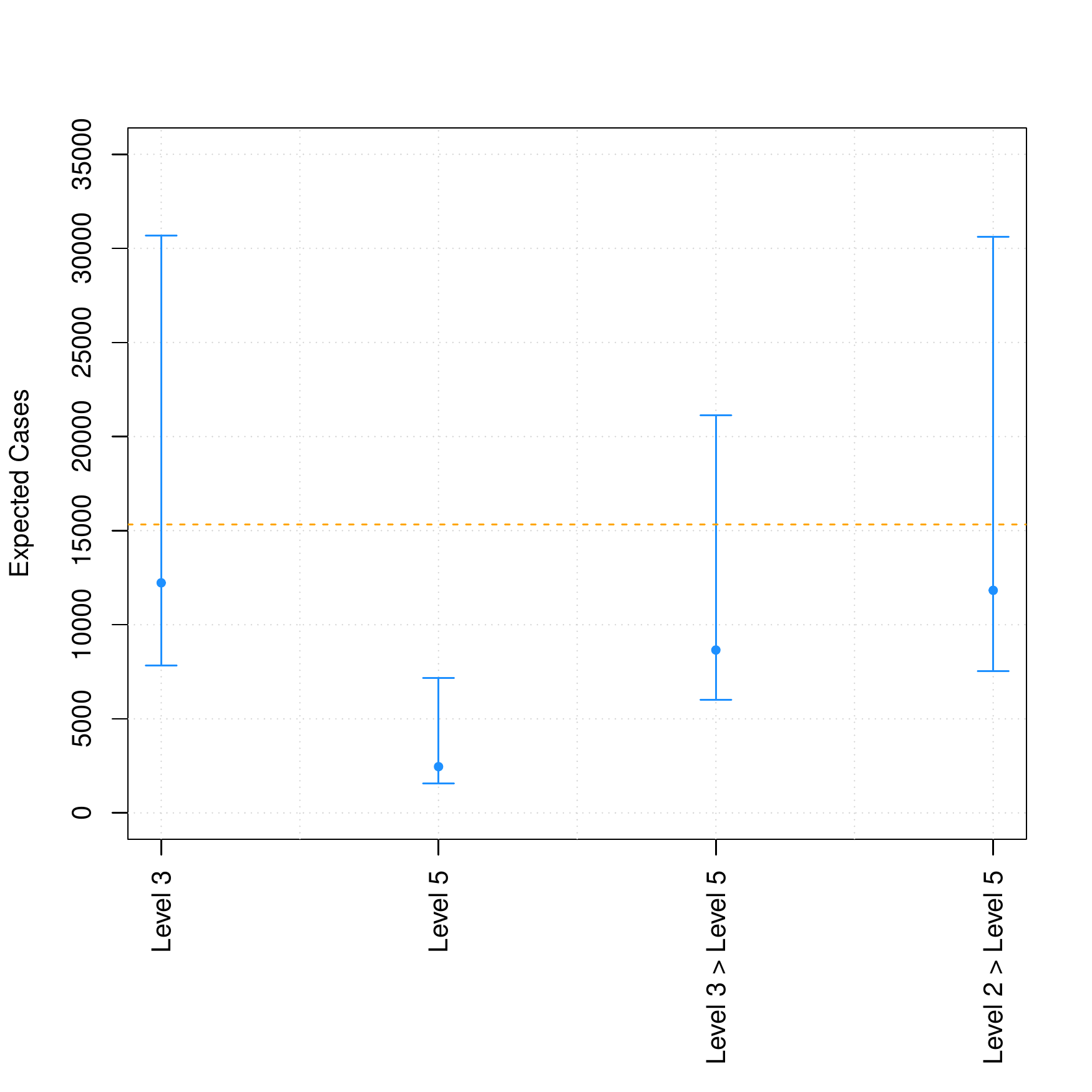}
	\caption{Estimated total cases for the 8 week period between 1st February 2021 and 30th March 2021.}
	\label{fig::projected_cases}
\end{figure}

\begin{figure}[ht]
	\centering
	\includegraphics[scale=0.57]{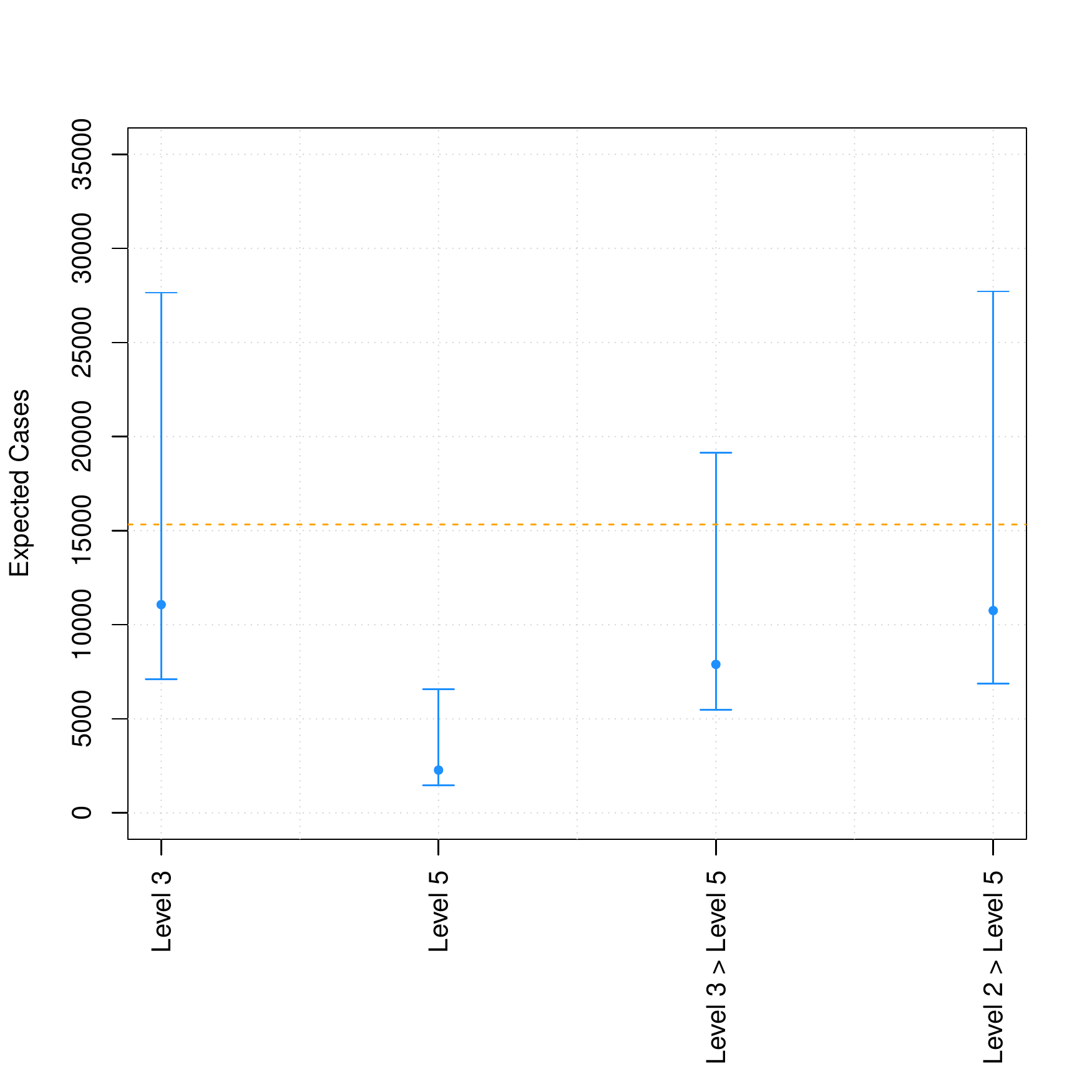}
	\caption{Estimated total cases for the 8 week period between 1st February 2021 and 30th March 2021 where the contacts for everyone aged 70 and over were set to zero after 1st December 2020.}
	\label{fig::projected_cases_isolated}
\end{figure}


\end{document}